\documentclass[a4paper,fleqn]{cas-dc}
\usepackage[numbers,sort&compress]{natbib}

\def\tsc#1{\csdef{#1}{\textsc{\lowercase{#1}}\xspace}}
\tsc{WGM}
\tsc{QE}
\tsc{EP}
\tsc{PMS}
\tsc{BEC}
\tsc{DE}
\begin{document}
\let\WriteBookmarks\relax
\def\floatpagepagefraction{1}
\def\textpagefraction{.001}
\shorttitle{Angle-Dependent Magnetoresistance  in Kagome Dirac Semimetal}
\shortauthors{Susanta et~al.}
\title [mode = title]{Extremely Large and Angle-Dependent Magnetoresistance in Kagome Dirac Semimetal RFe$_6$Sn$_6$ (R=Ho, Dy)}
\author[1,3]{Susanta Ghosh}
\credit{Conceptualization, Methodology, Validation, Formal analysis, Investigation, Writing original draft, Writing review \& editing}
\affiliation[1]{organization={S. N. Bose National Centre for Basic Sciences},
    addressline={Salt Lake, JD Block, Sector III, Bidhannagar},
    city={Kolkata},
    postcode={700106},
    state={West Bengal},
    country={India}}
\author[1,3]{Achintya Low}
\credit{Conceptualization, Validation, Writing review \& editing}
\author[2]{Nayana Devaraj}
\credit{Conceptualization, Validation,Investigation, Writing review \& editing}
\affiliation[2]{organization={Solid State and Structural Chemistry Unit},
    addressline={Indian Institute of Science},
    city={Bangalore},
    postcode={560012},
    state={Karnataka},
    country={India}}
\affiliation[3]{These authors contributed equally}
\author[1]{Susmita Changdar}
\credit{Writing review \& editing}
\author[2]{Awadhesh Narayan}
\credit{Conceptualization, Validation, Resources, Supervision, Project administration, Funding acquisition, Writing review \&
editing}
\author[1] {Setti Thirupathaiah}[orcid=0000-0003-1258-0981]
\cormark[1]
\ead{setti@bose.res.in}
\ead[URL]{www.qmat.in}
\credit{Conceptualization, Validation, Resources, Supervision, Project administration, Funding acquisition, Writing review \&
editing}

\begin{abstract}
We report on the electronic, magnetic, and magneto-transport properties  of  Fe-based kagome Dirac system, RFe$_6$Sn$_6$ (R = Ho, Dy). Magnetic properties study reveals  an antiferromagnetic order with N$\acute{e}$el temperature of $T_N \approx$ 570 K. Additionally,  a weak ferromagnetic order emerge at low temperatures. Magnetotransport measurements demonstrate  an extremely large magnetoresistance (XMR) reaching as high as $3\times 10^{3} \%$ for HoFe$_6$Sn$_6$ and  $ 1\times 10^{3} \%$ for DyFe$_6$Sn$_6$ when measured at 2 K with 9 T of magnetic field.  The semi-classical two-band model fitting of the Hall conductivity  reveals nearly perfect electron-hole compensation and high carrier mobility, which leads to XMR behaviour in these system. Further, we identify large magnetoresistance anisotropy for the magnetic fields applied in different crystallographic orientations. In addition,  considerable modification in the angle-dependent magnetoresistance (ADMR) pattern has been noticed between 2 and 50 K, indicating temperature-dependent changes in the Fermi surface topology of these systems.

 \end{abstract}

\begin{keywords}
Dirac Semimetal\\
Magnetic Materials\\
Magnetoresistance\\
Topological Hall effect\\
Angle-dependent Magnetoresistance
\end{keywords}

\maketitle

\section{Introduction}

Magnetoresistance (MR) is the change in the longitudinal electrical resistivity under an external magnetic field ($H$), which is defined as $ MR = [\rho(H) - \rho(0)]/\rho(0)$. Here,  $\rho(H)$ and $\rho(0)$ are the resistivity measured with and without magnetic field, respectively. Extremely large MR in magnetic materials is a rare phenomena~\cite{Heaps1939, Parkin1990, Niu2021} but has potential technological applications in spintronics~\cite{Ramirez1997, Yang2021, Vitayaya2024}. Therefore, searching for systems with extremely large magnetoresistance is a key research focus for physicists and materials scientists. Conventional non-magnetic metals generally display positive magnetoresistance (MR)~\cite{Pippard1989, Blundell2001,  Niu2021}, while ferromagnetic materials tend to exhibit negative MR~\cite{Heaps1939, Blundell2001,  Niu2021}. In most of the bulk magnetic systems, the MR (\%) values are relatively smaller and are typically below 5 \%~\cite{Heaps1939, Nickel1995, Blundell2001}, while the giant magnetoresistance (GMR) is found in low dimensional systems such as in thin films or multilayered structures~\cite{Tsymbal2001,Niu2021}. Also,  the colossal magnetoresistance (CMR) is mostly observed in the manganese-based perovskite oxides~\cite{Ramirez1997,Niu2021}. On the other hand, the tunneling magnetoresistance (TMR) is observed in magnetic tunnel junctions (MTJs) produced by sandwiching a thin insulating barrier between two ferromagnetic layers~\cite{Inoue1996}. Recently, an extremely large MR (XMR) of the order of $10^3\% - 10^8\%$  has been found in many nonmagnetic topological materials~\cite{Ali2014,Han2017,Gao2017,Okawa2018,Shekhar2015,Mondal2020}, triggering a great deal of research attention in this direction.

Though several topological systems have demonstrated the XMR in their bulk phase~\cite{Han2017,Gao2017,Okawa2018,Shekhar2015}, in the topological kagome systems it is very rarely observed. To the best of our knowledge, Ni$_3$In$_2$S$_2$ is the only kagome system that has been recently reported for the XMR behaviour~\cite{Zhang2024}. The kagome lattice, made up of geometrically frustrated corner-sharing triangles, has gained much attention recently due to their distinctive characteristics, including the van Hove singularities, flat bands, and Dirac points in the electronic band structure. Numerous kagome materials have been found to exhibit several interesting physical properties. For instance, the Co-based kagome system Co$_3$Sn$_2$S$_2$ shows giant anomalous Hall effect (AHE), Chiral anomaly, and Weyl fermions~\cite{Morali2019,Wang2018}. Despite being an antiferromagnet, the Mn-based kagome system Mn$_3$X (X=Sn, Ge) shows surprisingly large anomalous Hall effect, anomalous Nernst effect (ANE), and Weyl fermions~\cite{Nakatsuji2015,Nayak2016,Kuebler2018,Low2022,Changdar2023, Ghosh2023}. Whereas, the vanadium based kagome systems AV$_3$Sb$_5$ (A=K, Rb, Cs) show topological superconductivity~\cite{Jiang2021,Hu2022}.

In this regard, the RM$_6$X$_6$ (R = rare earth, M = transition metals, X = Sn, Ge) family of kagome systems have become a hot research topic due to their intriguing combination of electronic and magnetic properties. Within the RM$_6$X$_6$ kagome family, the researchers have so far focused on exploring the electronic and magnetic properties of the manganese (Mn)-based RMn$_6$X$_6$~\cite{Ma2021,Dhakal2021,Gao2021,Zhou2023,Xu2021, Kabir2022, Liu2023a,  Low2024} and the vanadium (V)-based RV$_6$X$_6$ ~\cite{Peng2021,Zhang2022,Lee2022,Arachchige2022} subfamilies but not much on the iron (Fe)-based RFe$_6$Sn$_6$ kagome family of systems.  The only experimental report available on the Fe-based YFe$_6$Sn$_6$ system suggests a low magnetoresistance (MR)~\cite{Liu2023}. Therefore, in this manuscript we systematically studied the magnetic and magnetotransport properties of the Fe-based RFe$_6$Sn$_6$ (R = Ho, Dy) kagome systems.

\begin{figure*}[t]
\centering
	\includegraphics[width=0.8\linewidth]{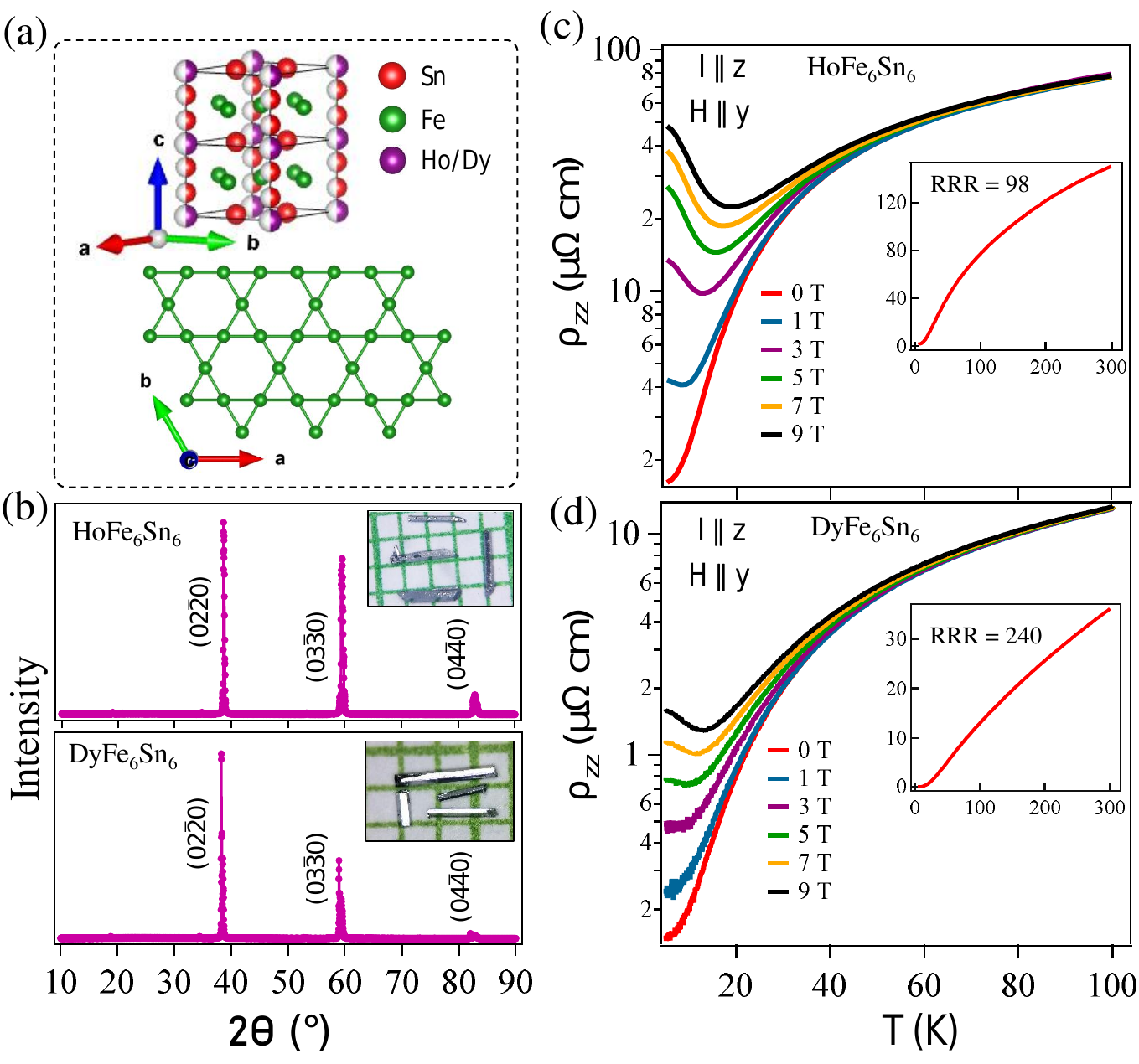}
	\caption{(a) Schematics of RFe$_6$Sn$_6$ crystal structure. (b) Powder X-ray diffraction patterns taken on the single crystal of HoFe$_6$Sn$_6$ and DyFe$_6$Sn$_6$. Insets in (b) show  the photographic image of HoFe$_6$Sn$_6$ and DyFe$_6$Sn$_6$ single crystal. Temperature dependence of longitudinal resistivity ($\rho_{zz}$), measured at different transverse magnetic fields ($H\parallel y$) for HoFe$_6$Sn$_6$ (c) and DyFe$_6$Sn$_6$ (d). Inset of (c) and (d) show the temperature dependence of resistivity measured without magnetic field for HoFe$_6$Sn$_6$ and  DyFe$_6$Sn$_6$, respectively.}
	\label{fig1}
\end{figure*}

\section{Experimental and Computational details}
Single crystals of RFe$_6$Sn$_6$ (R = Ho, Dy) were grown with Sn as flux using the high temperature muffle-furnace. The starting elements of high purity  $R$ (Ho, Dy) ingot (99.9\%, Alfa Aesar), Fe powder (99.99\%, Stren Chemical), and Sn shorts (99.995\%, Alfa Aesar) were mixed in the molar ratio of $1 : 6 : 30$. Next, an evacuated quartz ampoule was used to seal the alumina crucible containing the mixture.
The ampoule was then heated to $1000^{o}C$ at a rate of $100^{o}C/hr$ and held there for 18 hours, and cooled the molten mixture down to $600^{o}C$ at a rate of $5^{o} C/hr$. After annealing for another five days at $600^{o}C$, the ampoule was promptly transferred to a centrifuge to separate the as-grown single crystals from Sn-flux.
 In this approach, we obtained many hexagonal rod-shaped RFe$_6$Sn$_6$ (R = Ho, Dy) single crystals with a typical size of $(2 \times 0.3 \times 0.2)$ $mm^3$ as depicted in the inset of Fig.~\ref{fig1}(b).

The phase purity and crystal structure of the as-grown single crystals were analyzed using the powder X-ray diffraction (XRD, 9 KW, Rigaku Smart Lab) and single crystal XRD (SXRD, SuperNova, Rigaku) techniques with Cu-K$_\alpha$ radiation ($\lambda$ = 1.5406 \AA). The elemental composition of the as-grown single crystals were identified using energy-dispersive X-ray analysis (EDAX) coupled with scanning electron microscope (SEM, Quanta 250 FEG). The exact chemical composition were found to be the as-grown samples are found to be Ho$_{0.85}$Fe$_{6.12}$Sn$_{6.04}$ and Dy$_{1.16}$Fe$_{5.82}$Sn$_6$ using EDAX. For simplicity, we shall denote them by their nominal compositions of HoFe$_6$Sn$_6$ and DyFe$_6$Sn$_6$, wherever applicable.  Electrical transport, Hall effect, and magnetic properties were conducted using a 9 Tesla physical property measurement system (PPMS, Dynacool, Quantum Design) within the temperature range of 2-300 K. High temperature magnetic measurements up to 750 K were carried out using the VSM-oven option in the PPMS. Electrical transport and Hall effect properties were measured using the four-probe technique. For this, silver epoxy (EPO-TEK H20E) was used to attach copper leads to the sample.

Density functional theory (DFT) calculations were conducted using the Quantum ESPRESSO package~\cite{Giannozzi2009}. We employed ultrasoft pseudopotentials and  Perdew-Burke-Ernzerhof (PBE) functional within the generalized gradient approximation (GGA)~\cite{Perdew1997}. The kinetic energy cutoff for the wavefunction was at 140 Ry. To determine the equilibrium lattice parameters of HoFe$_6$Sn$_6$ and DyFe$_6$Sn$_6$, variable-cell relaxation calculations were executed with a $k$-grid of $6 \times 6 \times 4$. For the density of states (DOS) calculations, a denser $12 \times 12 \times 8$ $k$-grid was utilized. The convergence threshold for the self-consistent field was taken as 10$^{-9}$ Ry.

\section{Results and Discussions}

Fig.~\ref{fig1}(a) schematically shows the primitive unit cell of the hexagonal crystal structure of RFe$_6$Sn$_6$(top) and the crystal structure projected onto the $ab$-plane (bottom). From Fig.~\ref{fig1}(a), we can  see the $ab$-plane kogome lattice formed solely by the Fe-atoms. The XRD patterns obtained on the rod-shaped single crystals of RFe$_6$Sn$_6$ are shown in Fig.~\ref{fig1}(b), displaying the (0 2 $\bar{2}$ 0) Bragg's reflections, implying that the $c$-axis is the crystal growth axis. We further measured powder XRD on the crushed single crystals of RFe$_6$Sn$_6$ and performed Rietveld refinement using the Fullprof software (see Fig.~S1 in supplementary information). Rietveld refinement confirms that HoFe$_6$Sn$_6$ and DyFe$_6$Sn$_6$ crystallize into the hexagonal YCo$_6$Ge$_6$-type structure with a space group of $P6/mmm$ (191). The refined lattice parameters are $a = b = $5.3542 (5) \AA,  $c =$   4.4493(4) \AA,  $\alpha$ = $\beta$ = 90$^o$, $\gamma$ = 120$^o$ for HoFe$_6$Sn$_6$ and $a = b =$  5.3549 (5) \AA,  $c =$  4.4518 (4) \AA,  $\alpha$ = $\beta$ = 90$^o$, $\gamma$ = 120$^o$ for DyFe$_6$Sn$_6$. The hexagonal crystal structure is further confirmed by the single crystal XRD (SCXRD) (see Fig.~S2 in the supplemental information).

\begin{figure*}[b]
\centering
	\includegraphics[width=0.95\linewidth]{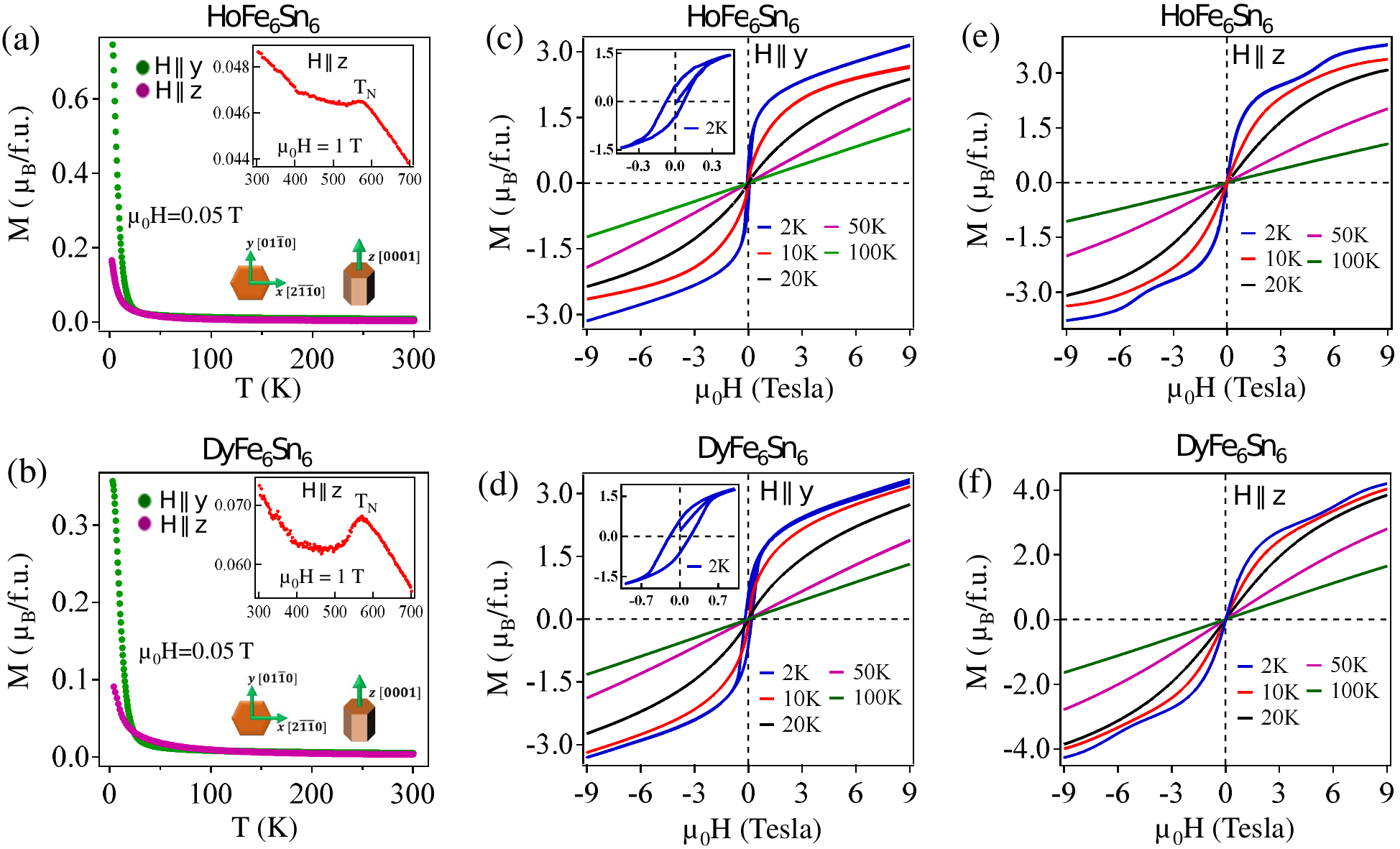}
	\caption{ Temperature dependent magnetization [$M(T)$],  measured for both $H\parallel z$ and $H\parallel y$ on the single crystals of (a) HoFe$_6$Sn$_6$ and (b) DyFe$_6$Sn$_6$. The inset of (a) and (b) shows the high temperature $M(T)$ data of HoFe$_6$Sn$_6$ and  DyFe$_6$Sn$_6$,  respectively. (c) and (d) show the magnetization isotherms [$M(H)$] measured at different temperatures for $H\parallel y$ from HoFe$_6$Sn$_6$ and  DyFe$_6$Sn$_6$, respectively. The inset of (c) and (d) shows the zoomed-in $M(H)$ at 2 K of HoFe$_6$Sn$_6$ and  DyFe$_6$Sn$_6$, respectively. Similarly, (e) and (f) show $M(H)$ data, measured at different temperatures for $H\parallel z$ from HoFe$_6$Sn$_6$ and  DyFe$_6$Sn$_6$, respectively.}
	\label{fig2}
\end{figure*}

 \begin{figure*}[hb]
 \centering
	\includegraphics[width=0.95\linewidth]{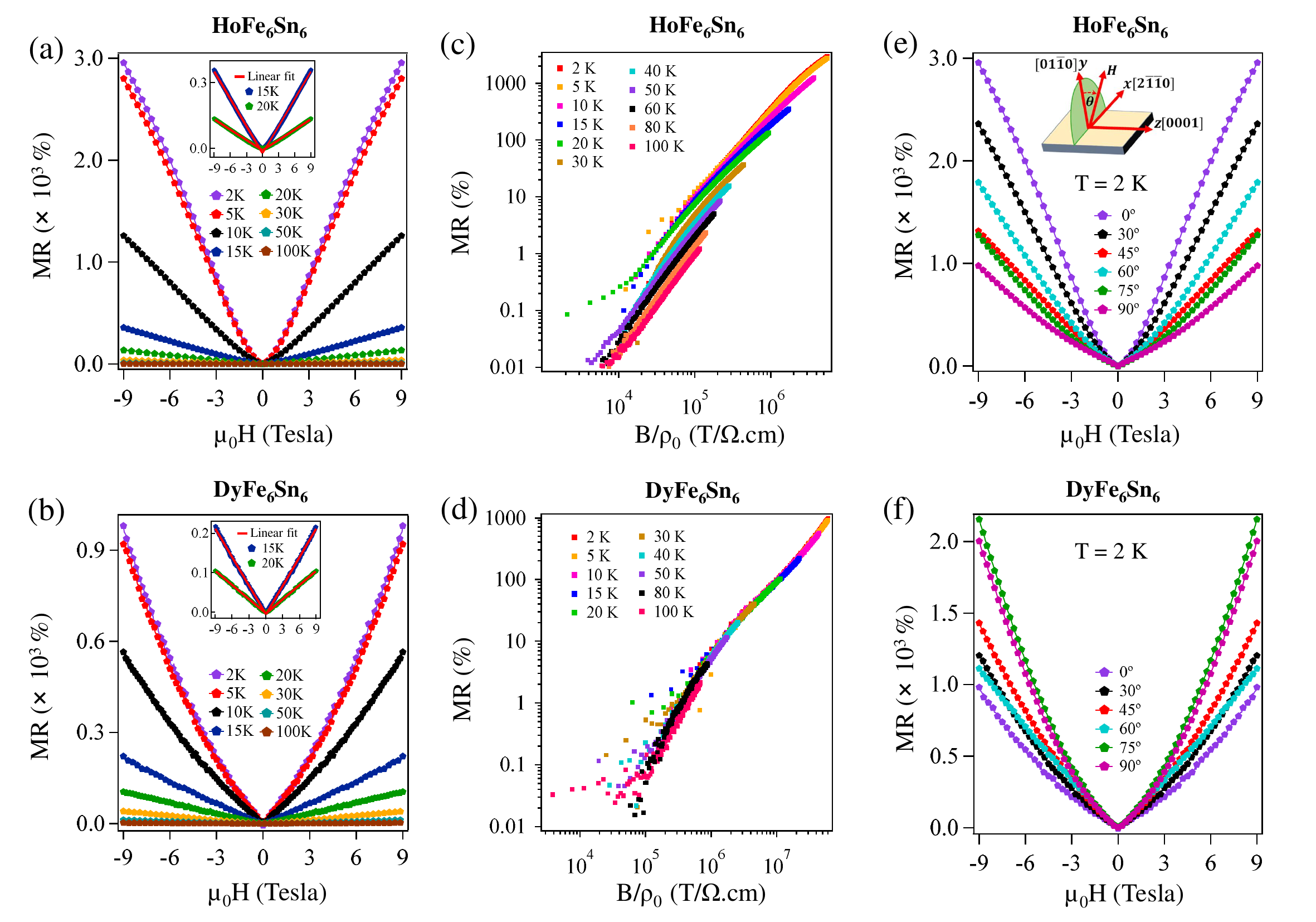}
	\caption{Magnetoresistance (MR) as a function of magnetic field is plotted for different temperatures from (a) HoFe$_6$Sn$_6$ and (b) DyFe$_6$Sn$_6$. The insets of (a) and (b) show MR data taken 15 and 20 K MR overlapped with liner fit from HoFe$_6$Sn$_6$ and DyFe$_6$Sn$_6$,  respectively. Scaling analysis of MR data using Kohler's rule for various temperatures  from (c) HoFe$_6$Sn$_6$ and (d) DyFe$_6$Sn$_6$.
MR as a function of field is measured at different field angles ($\theta$) when rotated in the $xy$-plane from (e) HoFe$_6$Sn$_6$ and (f) DyFe$_6$Sn$_6$. The inset in (e) illustrates measurement geometry.}
	\label{fig3}
\end{figure*}

Figs.~\ref{fig1}(c) and ~\ref{fig1}(d) exhibit temperature-dependent longitudinal electrical resistivity ($\rho_{zz}$) of HoFe$_6$Sn$_6$ and DyFe$_6$Sn$_6$, respectively, measured under various magnetic fields for $H\parallel y$ direction.  Inset of Figs.~\ref{fig1}(c) and ~\ref{fig1}(d) depict temperature-dependent longitudinal electrical resistivity with zero magnetic field from respective samples,  demonstrating a metallic behaviour throughout the measured temperature range. The residual resistivity ratio [$RRR = \rho_{zz}(300K) / \rho_{zz}(2K)$] is  about 98 for HoFe$_6$Sn$_6$ and 240 for DyFe$_6$Sn$_6$, suggesting good quality of the studied single crystals. As can be seen from Figs.~\ref{fig1}(c) and ~\ref{fig1}(d), the low temperature resistivity [$\rho_{zz}$ (T)] increases significantly with increasing field, demonstrating a significant magnetoresistance (MR) in the low temperature region.

Next, to explore the magnetic properties, temperature dependent magnetization [$M(T)$] and isothermal magnetization  [$M(H)$] at different temperatures are performed for $H\parallel z$ and $H\parallel y$ directions . The top panels of Fig.~\ref{fig2} show the magnetization data of HoFe$_6$Sn$_6$, whereas the bottom panels show the magnetization data of DyFe$_6$Sn$_6$. From high temperature magnetisation data, we observe an antiferromagnetic transition at around 570 K for both HoFe$_6$Sn$_6$ and DyFe$_6$Sn$_6$ single crystals [see inset in Figs.~\ref{fig2}(a) and ~\ref{fig2}(b)]. The magnetisation increases sharply at low temperatures, as seen in Fig.~\ref{fig2}(a) and Fig.~\ref{fig2}(b), suggesting a ferromagnetic type ordering at low temperatures. Using the first derivative of temperature-dependent magnetization data (not shown here), we identified the ferromagnetic ordering temperature of $T_C\approx$ 5.1 K from HoFe$_6$Sn$_6$ and 8.5 K from DyFe$_6$Sn$_6$.

\begin{figure*}[ht]
\centering
\includegraphics[width=1\linewidth, clip=true]{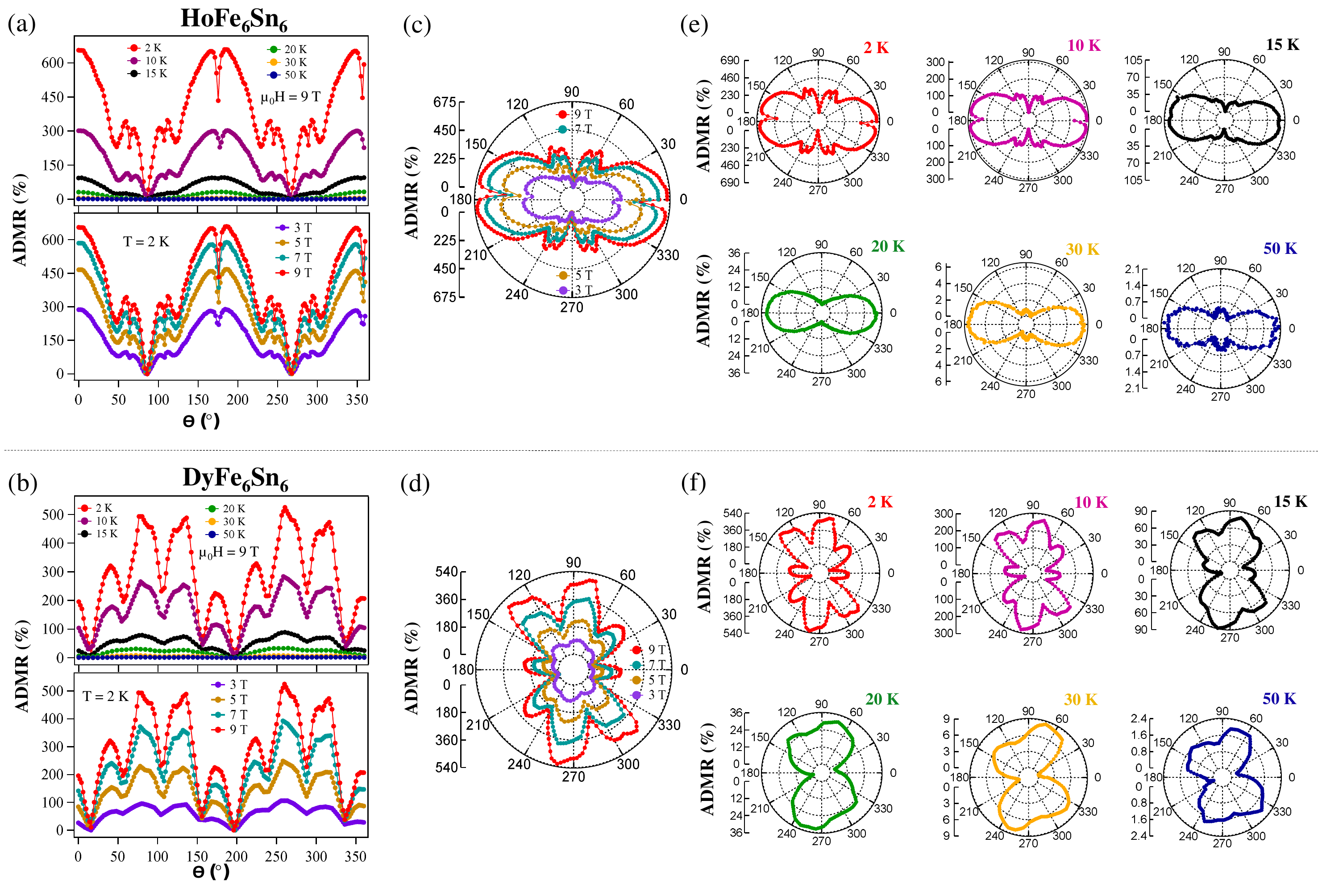}
\caption{Angular-dependent magnetoresistance (ADMR) measured for (a) HoFe$_6$Sn$_6$ and (b) DyFe$_6$Sn$_6$ by varying  the temperature at a fixed magnetic field of 9 T (top panels) and by varying the magnetic fields at a fixed temperature of 2 K (bottom panels). Current flows along the $z$-axis and the field is  rotated within the $xy$-plane as depicted in the inset of Fig.~\ref{fig3}(e). Polar plot of ADMR measured at 2 K by varying the field for (c) HoFe$_6$Sn$_6$ and (d) DyFe$_6$Sn$_6$.  Polar plot of ADMR measured at 9 T by varying the temperature for (e) HoFe$_6$Sn$_6$ and (f) DyFe$_6$Sn$_6$.}
\label{fig4}
\end{figure*}

\begin{figure*}[t]
\centering
\centerline{\includegraphics[width=0.95\linewidth, clip=true]{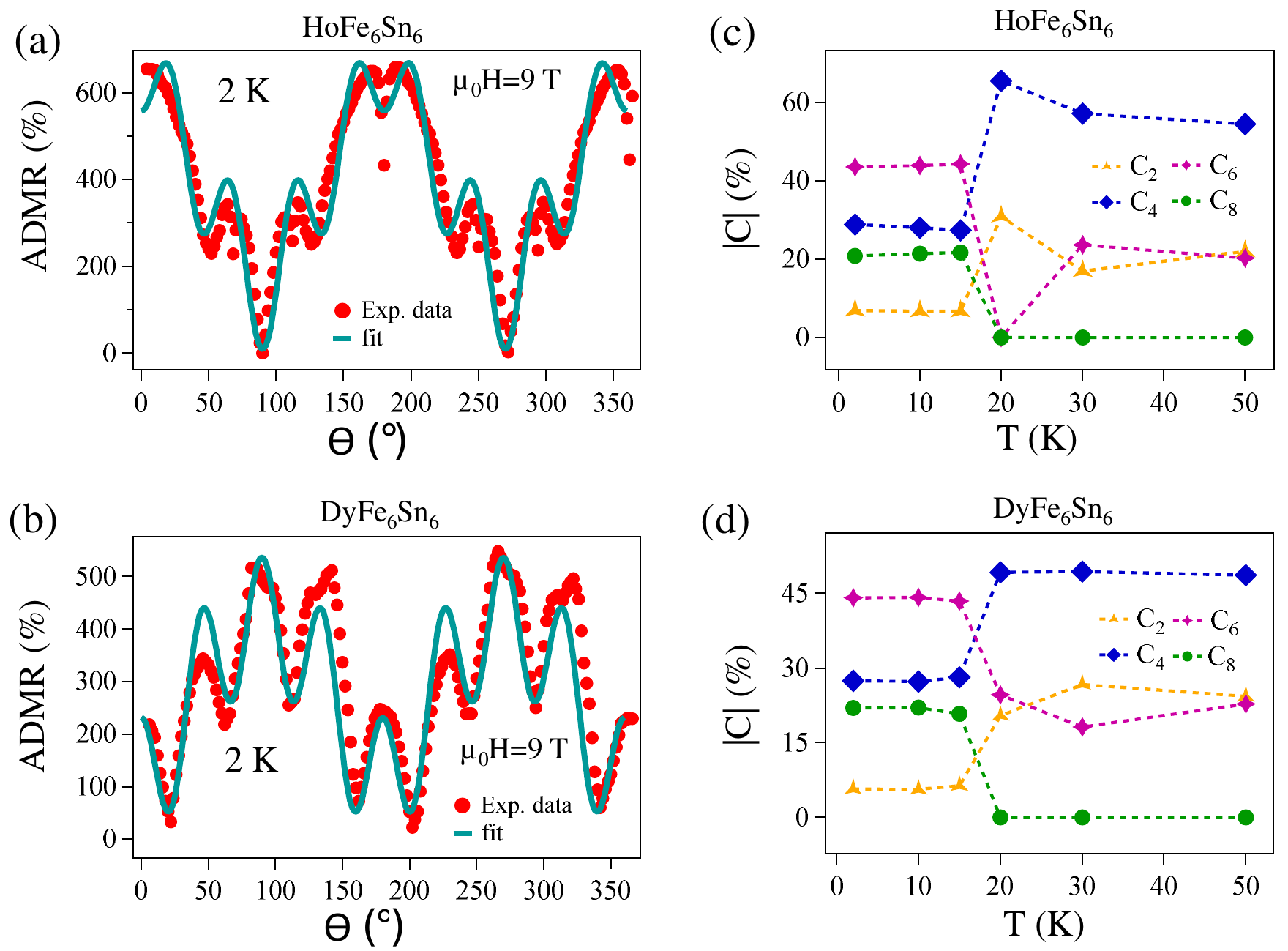}}
\caption{ADMR measured for (a) HoFe$_6$Sn$_6$ and (b) DyFe$_6$Sn$_6$ at 2 K and 9 T. The solid curves are fits of ADMR using the Eq.~\ref{Eq1} (see the text for more details). The normalized relative amplitudes (absolute values) C$_2$ (two-fold), C$_4$ (four-fold), C$_6$ (six-fold), and C$_8$ (eight-fold) components of the ADMR as a function of temperature are shown for (c) HoFe$_6$Sn$_6$ and (d) DyFe$_6$Sn$_6$. In (c) and (d), $|C_{2n}|=C_{2n}/\sum_{n=1}^{4} C_{2n}$.}
\label{fig5_1}
\end{figure*}

Figs.~\ref{fig2}(c) and ~\ref{fig2}(d) show the isothermal magnetization [$M(H)$], measured at various temperatures for $H\parallel y$ from HoFe$_6$Sn$_6$ and DyFe$_6$Sn$_6$ single crystals, respectively. The $M(H)$ curves at 2 K and 10 K exhibit a rapid increase in the low-field region, followed by a slower increase in the high-field region, without reaching saturation up to an applied field of 9 T. In agreement to the magnetization data which suggest ferromagnetic nature at low temperatures, we observe magnetic hysteresis for both systems when measured at 2 K [see the inset in Figs.~\ref{fig2}(c) and ~\ref{fig2}(d)]. Thus,  HoFe$_6$Sn$_6$ and DyFe$_6$Sn$_6$ show ferrimagnetic nature at low temperatures. However, $M(H)$ shows linear behaviour above 50 K due to melting of FM state and leaving behind the antiferromagnetic state alone.

Figs.~\ref{fig2}(e) and ~\ref{fig2}(f) show $M(H)$ data measured at various temperatures for $H\parallel z$ on HoFe$_6$Sn$_6$ and DyFe$_6$Sn$_6$ single crystals, respectively.  Interestingly, the $M(H)$ data measured at 2 K shows a $kink$ at around 5.25 T for  HoFe$_6$Sn$_6$ and 6.33 T for DyFe$_6$Sn$_6$,  persisting up to 10 K [see Fig. S8 in Supplemental information],  which indicates a field-induced metamagnetic state for $H\parallel z$. However, the metamagnetic state disappears at when measured at around 20 K for both HoFe$_6$Sn$_6$ and DyFe$_6$Sn$_6$. Such a low-temperature metamagnetic state was observed earlier too in systems like  DyV$_6$Sn$_6$, HoV$_6$Sn$_6$, RMn$_6$Ge$_6$ (R = Tb - Lu),  RAuGe (R = Dy, Ho, and Gd)~\cite{Zhou2023, Zeng2024, Kurumaji2024}. The metamagnetic state is commonly found in the rare-earth kagome based systems as the rare-earth ions such as Ho$^{3+}$ or Dy$^{3+}$ (in the present study) possess large, localized $4f$ magnetic moments that couple strongly to the crystalline electric field (CEF), giving rise to a pronounced single-ion magnetic anisotropy~\cite{Kurumaji2024}. The anisotropy constrains the spins to align along the preferred crystallographic directions. Further, when the magnetic field is applied along the hard axis of magnetization, the interplay between the magnetic anisotropy and the exchange interactions induces sudden spin flip/flop or canted spin magnetic moments, leading to the metamagnetic state at a suitable temperature and magnetic field. In contrast to the $M(H)$ data measured for $H\parallel y$, magnetic hysteresis is not found from both HoFe$_6$Sn$_6$ and DyFe$_6$Sn$_6$  for $H\parallel z$. Nevertheless, similar to $H\parallel y$, the systems become completely antiferromagnetic above 50 K for $H\parallel z$ as well.

Earlier,  in the case of Fe-based  RFe$_6$Ge$_6$ and RFe$_6$Sn$_6$ (R = rare earth) series of intermetallic compounds an independent magnetic behaviour of the rare-earth (R) and Fe sublattices was identified~\cite{Cadogan2001,Cadogan2001a,Cadogan2006}. In RFe$_6$Sn$_6$, while the Fe sublattice shows an antiferromagnetic ordering below the N$\acute{e}$el temperature of $T_N\approx$ 560 K, the rare-earth sublattice  orders  ferromagnetically at significantly lower temperatures. For R = Gd - Er, the ferromagnetic ordering temperature of rare-earth sublattice ranges from  3 K for ErFe$_6$Ge$_6$ to 45 K for GdFe$_6$Sn$_6$,  without affecting the Fe sublattice AFM ordering temperature. Our experimental findings also reveal a similar magnetic ordering pattern for the  HoFe$_6$Sn$_6$ and DyFe$_6$Sn$_6$  exhibiting an antiferromagnetic ordering at around 570 K whereas a ferromagnetic ordering is found at  5.1 K for HoFe$_6$Sn$_6$ and  8.5 K for DyFe$_6$Sn$_6$. These observations are in well agreement with previous reports on similar systems~\cite{Cadogan2001,Cadogan2001a,Cadogan2006}. Further, the estimated magnetic moment per Ho in HoFe$_6$Sn$_6$ is about 3.78 $\mu_B$ for $H\parallel z$ and 3.11 $\mu_B$ for $H \parallel y$. These values are substantially lower than the magnetic moments of free  $Ho^{3+}$ (10.0 $\mu_B$) ion. The same has been noticed from  DyFe$_6$Sn$_6$, that the magnetic moment per Dy is about 4.20 $\mu_B$ for H $\parallel z$ and 3.34 $\mu_B$ for $H\parallel y$, which is also substantially lower than the  magnetic moment of free $Dy^{3+}$ (10.6 $\mu_B$) ion. This observation clearly confirms that the magnetic moment of the measured systems is not fully polarised even at an external magnetic field of 9 T, resulting into the coexistence of both FM and AFM states in  Ho(Dy)Fe$_6$Sn$_6$ single crystals at low temperatures.

Next, coming to the important observations of this study, Figs.~\ref{fig3}(a) and ~\ref{fig3}(b) depict the transverse magnetoresistance (MR), i.e., the resistance measured under transverse electric and magnetic fields, at different sample temperatures for HoFe$_6$Sn$_6$ and DyFe$_6$Sn$_6$, respectively. Here, we measure MR with the current applied along $z$-axis and field applied along $y$-axis. We calculate MR using the relation  $MR(\%) = [{\rho_{zz} (T,\mu_0H) - \rho_{zz} (T,0)}]/{\rho_{zz} (T,0)}\times 100 \%$. In both systems, a positive non-saturating magnetoresistance was observed  at all measured temperatures with the field applied up to 9 T. At 2 K and 9 T, MR reaches  a maximum value of $ 3\times 10^{3} \%$ for HoFe$_6$Sn$_6$ and  $ 1\times 10^{3} \%$ for DyFe$_6$Sn$_6$. The value of MR decreases rapidly with increasing temperature and reaches $1 \%$ for HoFe$_6$Sn$_6$ and $2 \% $ for DyFe$_6$Sn$_6$ at 100 K and 9 T. This is surprising as the magnetic bulk intermetallic systems usually do not show large MR. Also, let us emphasize here that this is first study showing such as extremely large magnetoresistance from the magnetic Fe-based kagome systems. The only other kagome system reported to exhibit XMR is M$_3$In$_2$S$_2$ (M = Ni,Co) kagome systems~\cite{Zhang2024, Lv2024}.

To understand the nature magnetoresistance, we fitted the field dependent MR with  equation, MR $\varpropto$ $B^m$ [see Fig. S3 in the supplementary information] ~\cite{Singha2017}. From the fittings, we  estimate $m\simeq$ 1.02 for HoFe$_6$Sn$_6$ and  1.45 for DyFe$_6$Sn$_6$ at 2 K.  At, 15  and 20 K, the MR follows linear field dependence ($m=$1) for both crystals as shown in the inset of Figs.~\ref{fig3}(a) and ~\ref{fig3}(b). At 100 K, we estimated  $m\simeq$ 1.74 for HoFe$_6$Sn$_6$ and   1.79 for DyFe$_6$Sn$_6$ [see Fig. S3 in the supplemental information]. Conventional MR generally changes quadratically with field, but in our case we find that the MR dependence on field changes with temperature. To further confirm this phenomenon, we performed scaling analysis using the Kohler's rule and plotted the MR curves as a function of $B/\rho_0$ (Kohler's plot) measured at different temperatures as shown in Figs.~\ref{fig3}(c) and ~\ref{fig3}(d). From the Kohler's plots it is evident that the MR curves do not collapse onto a single universal curve for all measured temperatures, indicating the break down of Kohler's rule. The violation of Kohler's rule suggests a single scattering mechanism is not applicable to explain the MR in the studied systems. Therefore, different temperature regions have distinct scattering mechanism~\cite{Narayanan2015,Zhu2022,Zhang2024}.

Further, we measured the field-dependent MR at 2 K by applying the magnetic field along different crystallographic axis while fixing the current direction. The inset of  Fig.~\ref{fig3}(e) schematically shows the angle-dependent MR (ADMR) configuration, with the current along the $z$ [0001] axis and the magnetic field direction varies in the $xy$-plane.  As can be seen from Fig.~\ref{fig3}(e), in HoFe$_6$Sn$_6$, the MR gradually decreases from  $3\times 10^{3}$ $\%$  to  $1\times 10^{3}$ $\%$ as the angle $\theta$ increases from $0^{\circ}$ to $90^{\circ}$. Conversely, the MR in DyFe$_6$Sn$_6$ increases from $1\times 10^{3}$ $\%$ to ${2}\times 10^{3}$ $\%$ as we increase  $\theta$ from $0^{\circ}$ to $90^{\circ}$ and reaching maximum $ {2.2}\times 10^{3}$ at $\theta$ = $75^{\circ}$ [Fig.~\ref{fig3}(f)]. Thus, we can see that the MR is highly crystallographic axis dependent. To further investigate the directional dependent MR, we fit the data using a power law function (MR $\varpropto$ $B^m$) at different angles as shown in the Fig.~S4 of the supplemental information. From the fittings, we estimate $m\simeq$ 1.02, 1 and 1.36 at the angles $0^{\circ}$, $45^{\circ}$ and $90^{\circ}$ for HoFe$_6$Sn$_6$ and  1.45, 1, and, 1.54 at the angles $0^{\circ}$, $60^{\circ}$ and $90^{\circ}$ for DyFe$_6$Sn$_6$, respectively. We find linear MR at 15  and 20 K for both the HoFe$_6$Sn$_6$ and DyFe$_6$Sn$_6$ systems at an angle of $0^{\circ}$.

For more insights on the angle-dependent magnetoresistance (ADMR), we measured MR at different temperatures under a fixed magnetic field of 9 T and by rotating the field direction with respect to the crystal axis as shown in Figs.~\ref{fig4}(a) and ~\ref{fig4}(b) for HoFe$_6$Sn$_6$ and DyFe$_6$Sn$_6$, respectively.  For a given field and temperature the ADMR is calculated using the formula,  $ADMR(\%)$ = $\frac{\rho_{zz} (\theta)  -  \rho_{zz} (\theta_{min})}{\rho_{zz} (\theta_{min})} \times 100\%$. $\theta_{min}$ is  about $86^{\circ}$ for HoFe$_6$Sn$_6$, whereas $\theta_{min}$ is about $196^{\circ}$ for DyFe$_6$Sn$_6$. In Fig.~\ref{fig4}(a), the ADMR measured at 2 K shows a total of 12 local maxima and 12 local minima, resulting in a total of 12 lobes for HoFe$_6$Sn$_6$. This pattern remains unchanged up to 10 K. However, as the temperature increases further the ADMR pattern evolves by decreasing the  number of lobes. Similarly, in Fig.~\ref{fig4}(b), the ADMR at 2 K for DyFe$_6$Sn$_6$ displays 8 local maxima and 8 local minima, forming the ADMR pattern with 8 lobes. As the temperature increases, the number of lobes decreases similar to HoFe$_6$Sn$_6$. From Figs.~\ref{fig4}(a) and~\ref{fig4}(b), it is evident that the MR is highly sensitive to field angle, indicating the presence of strong anisotropic MR in these systems.

For a better visualization, we plotted the ADMR in polar graphs as depicted in Figs.~\ref{fig4}(c)-~\ref{fig4}(f). Figs.~\ref{fig4}(c) and ~\ref{fig4}(d) show the polar-plot of ADMR measured at 2 K under different magnetic fields for HoFe$_6$Sn$_6$ and DyFe$_6$Sn$_6$, respectively.  Figs.~\ref{fig4}(e) and~\ref{fig4}(f) show the polar-plot of ADMR measured under an applied field of 9 T at various sample temperatures for HoFe$_6$Sn$_6$ and DyFe$_6$Sn$_6$, respectively. Further, from Fig.~\ref{fig4}(c),  we can see a butterfly-like ADMR pattern for HoFe$_6$Sn$_6$  at 2 K and the pattern remains almost constant, though the value of MR\% decreases with decreasing field. Similarly, ADMR pattern of  DyFe$_6$Sn$_6$ [see  Fig.~\ref{fig4}(d)] also looks butterfly-like but rotated by 90$^{\circ}$.  On the other hand, from Figs.~\ref{fig4}(e) and~\ref{fig4}(f), one can clearly see that the polar-plot of ADMR changes significantly with temperature. Importantly, the evolution of ADMR with temperature is different between HoFe$_6$Sn$_6$ and DyFe$_6$Sn$_6$. For instance, in the case of HoFe$_6$Sn$_6$ [see  Fig.~\ref{fig4}(e)], the butterfly-like pattern with 12 lobes are visible at 2 and 10 K. As the temperature rises, the ADMR pattern changes from a butterfly-like to a dumbbell-like at 20 K, with only two lobes visible. Further increasing the temperature,  at 30  and 50 K, an additional two lobes have reemerged at $90^{\circ}$ and $180^{\circ}$. In the case of DyFe$_6$Sn$_6$ [see  Fig.~\ref{fig4}(f)],  the butterfly pattern, with eight lobes, is visible up to 15 K. Between 20 and 50 K, only four lobes are visible.

Next, Figs.~\ref{fig5_1}(a) and ~\ref{fig5_1}(b) depict the ADMR measured at 2 K under the magnetic field of 9 T for HoFe$_6$Sn$_6$ and DyFe$_6$Sn$_6$, respectively.  The ADMR data is fitted by Eq.~\ref{Eq1} having contribution up to 8-fold symmetry. Although the fitting is not perfect, the Eq.~\ref{Eq1} can reasonably reproduce the experimental data. The fittings for higher temperature ADMR data are shown in the Fig.~S5 of the supplemental information.   The normalized relative amplitudes C$_2$ (two-fold), C$_4$ (four-fold), C$_6$ (six-fold), and C$_8$ (eight-fold) of ADMR are plotted as a function of temperature in Figs.~\ref{fig5_1}(c) and ~\ref{fig5_1}(d) for HoFe$_6$Sn$_6$ and DyFe$_6$Sn$_6$, respectively. From the temperature dependent relative amplitudes, we can notice that all symmetry components contribute to the ADMR up to 20 K. However, from 20 K, the eight-fold symmetry contribution to ADMR is totally suppressed.

 \begin{equation}
 \centering
  ADMR = C_0+ C_2~Cos^2 \theta + C_4~Cos^4 \theta +
  C_6~Cos^6 \theta+ C_8~Cos^8 \theta
  \label{Eq1}
 \end{equation}

where $C_0$ is a constant,  $C_2 ~Cos^2 \theta$,  $C_4 ~Cos^4 \theta$, $C_6 ~Cos^6 \theta$, and $C_8 ~Cos^8 \theta$ account for the two, four, six, and eight fold components, respectively. The constants $C_2$, $C_4$, $C_6$, and $C_8$  are the amplitudes.

\begin{figure*}[ht]
\centering
\includegraphics[width=1\linewidth, clip=true]{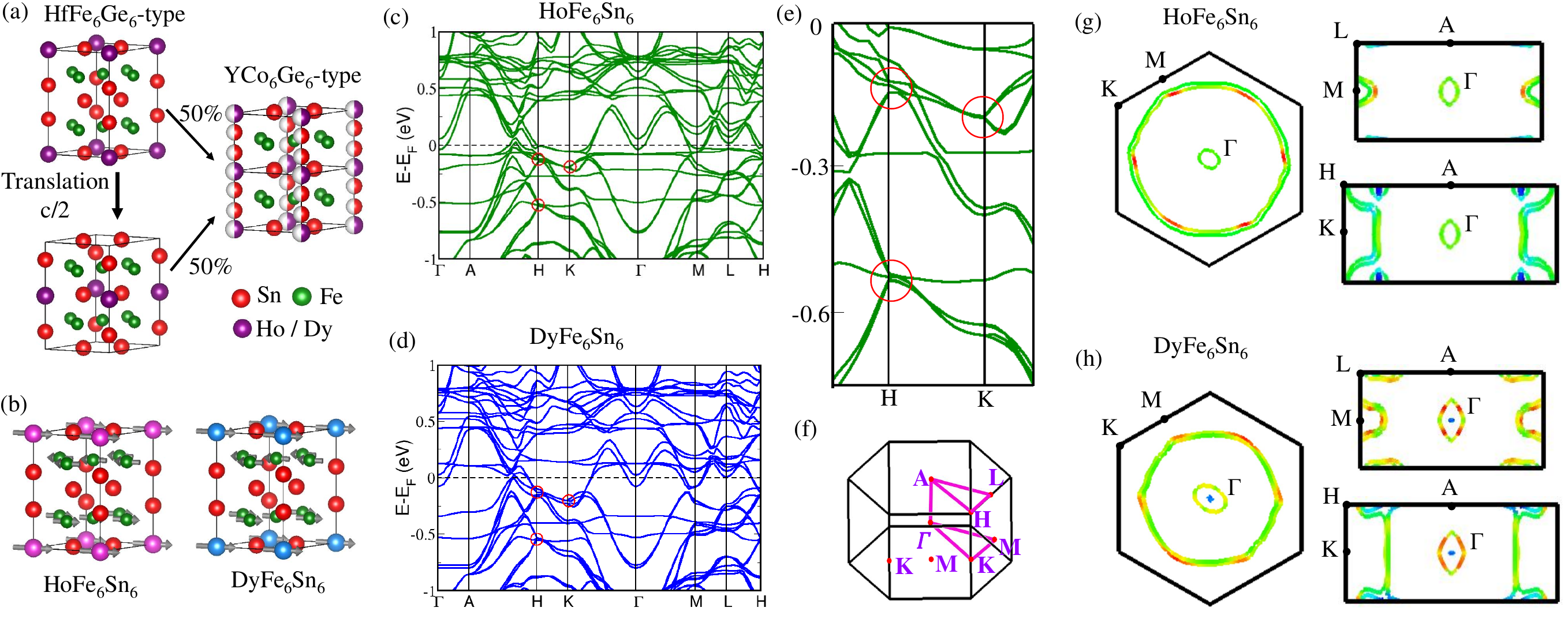}
\caption{(a) Schematic demonstration of two HfFe$_6$Ge$_6$-type crystal structures forming one partially filled YCo$_6$Ge$_6$. (b) Ground state magnetic structure of HoFe$_6$Sn$_6$ and  DyFe$_6$Sn$_6$,  derived from the DFT calculations. (c) and (d) Electronic band structures of HoFe$_6$Sn$_6$ and DyFe$_6$Sn$_6$, respectively, calculated without including spin-orbit coupling (SOC). (e) Zoomed-in image of the (c) highlighting the Dirac-like band dispersions near the Fermi level. (f) Hexagonal Brillouin zone with high symmetry points located.  (g) and (h) In-plane and out-of-plane Fermi contours of HoFe$_6$Sn$_6$ and DyFe$_6$Sn$_6$, respectively, calculated without including SOC.}
\label{fig6}
\end{figure*}

Several mechanisms exist in the literature explaining the manyfold symmetric ADMR, such as (i) magnetocrystalline anisotropy~\cite{Jin2012, Miao2021, Zhang2023}, (ii) spin scattering near the antiphase boundaries (APBs)~\cite{Ramos2008, Li2010, Li2010a}, (iii) exchange bias~\cite{Li2015}, (iv) relaxation time anisotropy~\cite{Dai2022, Lv2024}, (v) valleytronics~\cite{Zhou2020}, (vi) symmetry of the lattice~\cite{Hu2012, Lv2024}, or (vii) density of states modulation near the Fermi level (such as  Lifshitz transitions)~\cite{Dai2022,  Chen2023}. As can be seen from Figs.~\ref{fig4}(c) and ~\ref{fig4}(d), the ADMR is insensitive to the applied field. Therefore, the observed ADMR in HoFe$_6$Sn$_6$ and DyFe$_6$Sn$_6$ may not be originated from the magnetism. Further, the mean free path of the charge carriers calculated from Hall effect measurement data (shown later) at 2 K  is about $25.20$ $\mu m$ for HoFe$_6$Sn$_6$ and  $15.96$ $\mu m$ for DyFe$_6$Sn$_6$, which are much higher than the distance of $Fe-Fe$ (2.67 \AA), $Ho-Ho$ (5.35 \AA),  or $Dy-Dy$ (5.35 \AA) magnetic moments. Therefore, we can neglect the spin-charge scattering in these systems, allowing us to safely exclude the influence of magnetism on the observed ADMR as discussed in the points (i)-(iv). Also, we can safely neglect the valleytronic origin of ADMR as the vallies are highly sensitive to the orientation and magnitude of the applied magnetic ﬁeld~\cite{Zhu2018} which is not the case in the current study. The other possibilities of ADMR are the crystal symmetry  or the band structure near the Fermi level. As discussed above, the studied systems shows 2-, 4-, 6-, and 8-fold symmetry in the ADMR at low temperature ($<$ 20 K) and at higher temperatures the 8-fold component completely gets suppressed leaving only the 2-, 4-, and 6-fold symmetries. Particularly,  the 2-fold symmetry contribution is very small at low temperatures, and thus the total ADMR is dominated by the higher-fold symmetries. At higher temperatures,  except the 8-fold symmetry component,  the other symmetries significantly contribute to the total ADMR. Since the field is applied always perpendicular to the $z$-axis and is rotated within the $xy$-plane, the 6-fold symmetry contribution could be originated from the DOS modulation near the Fermi level and from the hexagonal crystal symmetry. On the other hand, the 2-fold and 4-fold symmetry contributions to the ADMR might have originated from the DOS modulation near the Fermi level~\cite{Dai2022,  Chen2023, Lv2024}.  This is because, in the momentum space, the charge carriers orbit around the Fermi surface cross-sections that are perpendicular to the magnetic field (B) direction~\cite{Wang2019}. As a result, the anisotropic shape of the Fermi surface leads to variation in the cyclotron mass and velocity, following the relation $\upsilon_k = \frac{1}{\hbar}\nabla_k\varepsilon_k$. This variation in the cyclotron mass and velocity,  has an effect on how easily the charge carrier trajectories bend under the magnetic field, influence the value of MR~\cite{Zhang2019,Wang2019}.

To better understand band structure contribution to the anisotropic MR in these systems, we performed the density functional theory (DFT) calculations. Performing DFT calculations on Ho(Dy)Fe$_6$Sn$_6$ (YCo$_6$Ge$_6$-type structure) system is quite challenging due to its disordered nature. As illustrated in Fig.~\ref{fig6}(a), the YCo$_6$Ge$_6$-type structure can be viewed as a combination of two HfFe$_6$Ge$_6$-type unit cells where one of them is translated by $c/2$~\cite{Weiland2020,Liu2023}. To address the challenges associated with the disordered and partially occupied YCo$_6$Ge$_6$-type structure, we used the ordered HfFe$_6$Ge$_6$-type structure for the DFT calculations. This approach aligns well with previous studies on the related systems~\cite{Liu2023,Huang2023}. Although the HfFe$_6$Ge$_6$-type and YCo$_6$Ge$_6$-type structures share the same space group ($P6/mmm$, 191), they differ in their atomic arrangement. Our DFT calculations suggest that in the ground state the Fe magnetic moments align antiferromagnetically, while the Ho(Dy) magnetic moments align ferromagnetically within the $ab$-plane as shown in Fig.~\ref{fig6}(b). This prediction is inline with our magnetization measurements which suggest that the easy magnetization axis lies on the $ab$-plane [see Fig.~\ref{fig2}].

 \begin{figure*}[ht]
 \centering
	\includegraphics[width=\linewidth]{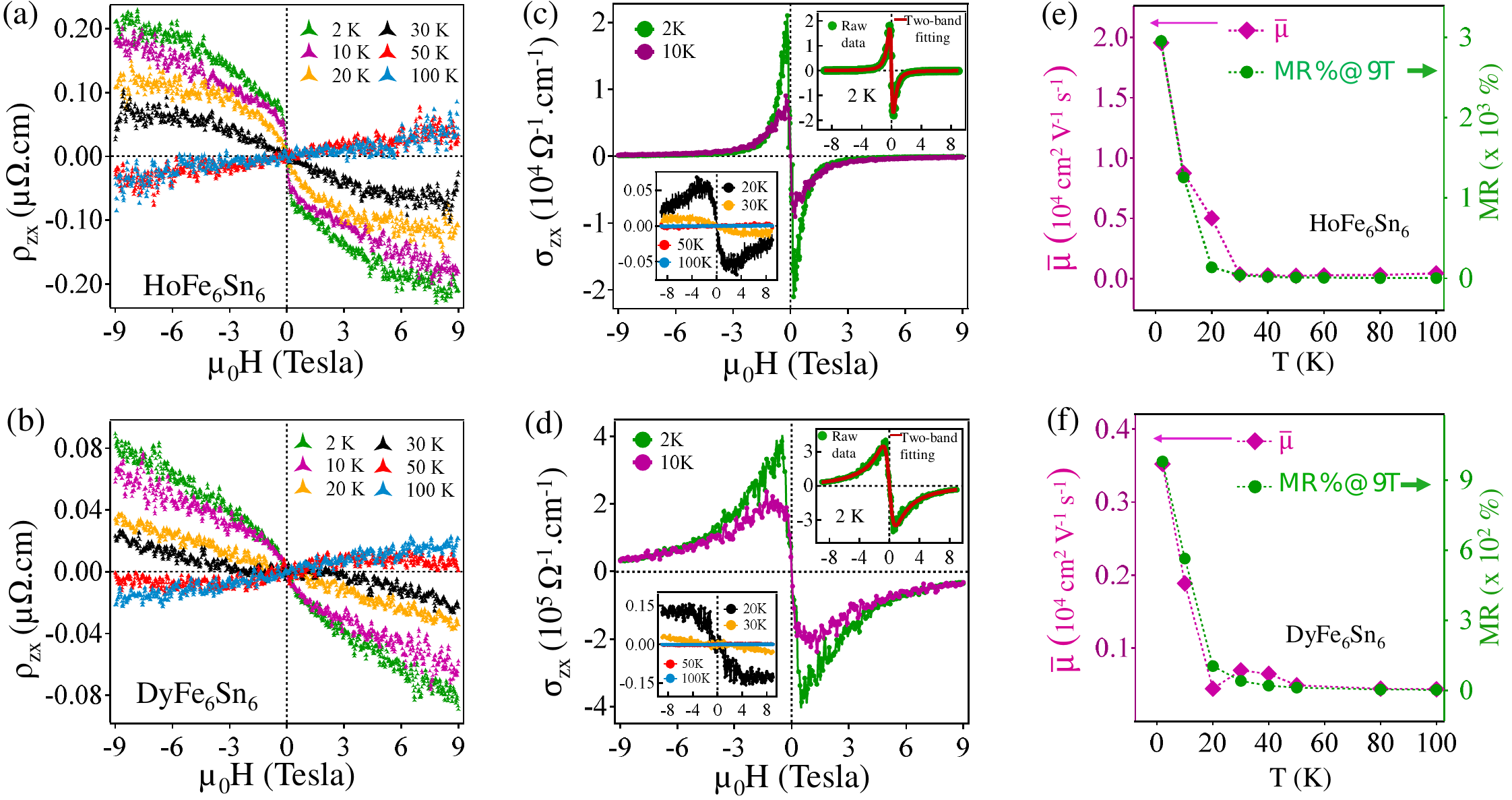}
	\caption{Field-dependent Hall resistivity ($\rho_{\it{zx}}$) measured at various temperatures for (a) HoFe$_6$Sn$_6$ and (b) DyFe$_6$Sn$_6$. Hall conductivity ($\sigma_{\it{zx}}$) derived from Hall resistivity ($\rho_{\it{zx}}$)  at 2 K and 10 K of (c) HoFe$_6$Sn$_6$ and (d) DyFe$_6$Sn$_6$. Bottom left inset of (c) and (d) are the Hall conductivity ($\sigma_{\it{zx}}$) derived at 20 and 30 K. Top right inset of (c) and (d) show two-band model fitting of  $\sigma_{\it{zx}}$ at 2 K. Mean carrier mobility (left axis), $\bar{\mu} = \sqrt{\mu_h \mu_e}$ and MR(\%) at 9 T (right axis) are plotted as a function of temperature for (e) HoFe$_6$Sn$_6$ and (f) DyFe$_6$Sn$_6$.}
	\label{fig5}
\end{figure*}

Figs.~\ref{fig6}(c) and ~\ref{fig6}(d) show the electronic band structure plotted for HoFe$_6$Sn$_6$ and DyFe$_6$Sn$_6$, respectively, without considering the spin-orbit coupling (SOC). The band dispersions of both systems qualitatively look similar. Both band structures exhibit several parabolic bands crossing the Fermi level, confirming the metallic nature of the studied systems. Interestingly, we notice several non-dispersive flat bands near the Fermi level mostly contributed by the Ho(Dy) $4f$ orbitals [see Fig.S7 in the supplemental information]~\cite{Peng2021,Guo2023,Xu2023}. Further, from a zoomed-in band structure shown in Fig.~\ref{fig6}(e), we observe several Dirac-like band crossings in the vicinity of the Fermi level marked by red-circles.  These band crossings are mainly originating from the Fe $3d$ orbitals. Next, Figs.~\ref{fig6}(g) and ~\ref{fig6}(h) show the Fermi maps of HoFe$_6$Sn$_6$ and DyFe$_6$Sn$_6$, respectively, from the in-plane and out-of-plane momentum space. From the in-plane ($\Gamma K M$) Fermi map, we mainly observe one small and two large Fermi sheets around $\Gamma$ point. On comparing the Fermi sheets with the electronic band dispersions shown in Fig.~\ref{fig6}(e), we can conclude that the two large Fermi contours are of hole-type and the smaller one is of electron-type.

Similarly, from the Fermi map taken in the $\Gamma M L A$ plane, we could observe two small electron-like highly dispersive out-of-plane Fermi sheets along the $M-L$ direction and another electron-type Fermi sheet dispersing along the $\Gamma-A$ direction. In addition, from the Fermi map taken in the $\Gamma K H A$ plane, we could observe two out-of-plane highly dispersive large electron-type Fermi sheets along the $K-H$ direction and two hole-like in-plane Fermi sheets along the $H-A$ direction. Since the applied field is perpendicular to the $z (k_z)$-axis, the Fermi sheets lying in the $\Gamma K H A$ and $\Gamma M L A$ planes contribute to the magnetoresistance. And can be seen from Figs.~\ref{fig6}(g) and ~\ref{fig6}(h), the Fermi sheets in these planes show 2- and 4-fold (nearly) symmetries. On rotating the magnetic field axis in the $xy$-plane, we could get the 6-fold modulation of the Fermi sheets of the $\Gamma K H A$ and $\Gamma M L A$ planes.  We note that the band structures presented here are calculated without spin-orbit coupling. Spin-orbit coupling calculations are very challenging and lead to convergence issues due to the presence of $4f$ states.

Next, as the Hall effect is sensitive to the topological band structure, we performed Hall effect measurements for both HoFe$_6$Sn$_6$ and DyFe$_6$Sn$_6$ systems.  The field-dependent Hall resistivity ($\rho_{zx}$) measured at different temperatures is shown in Figs.~\ref{fig5}(a) and ~\ref{fig5}(b) for HoFe$_6$Sn$_6$ and DyFe$_6$Sn$_6$, respectively. Here, $\rho_{\it{zx}}$ corresponds to the current applied along the $\it{z}$-direction, field applied along the $\it{y}$-direction, and the Hall voltage was measured along the $\it{x}$-direction.  From Fig.~\ref{fig5}(a), it is clear that the Hall resistivity is not linear with magnetic filed, suggesting more than one-type of charge carriers dominating the magnetotransport. Importantly, the slope of $\rho_{\it{zx}}$ $vs.$ H curve is positive below 30 K and negative for 50 and 100 K. We could not measure the Hall effect above 100 K due to high noise to signal ratio. Similar behaviour was also observed from DyFe$_6$Sn$_6$ as shown in Fig.~\ref{fig5}(b). Figs.~\ref{fig5}(c) and ~\ref{fig5}(d) show the  Hall conductivity ($\sigma_{\it{zx}}$) calculated from the Hall resistivity ($\rho_{\it{zx}}$) and longitudinal resistivity ($\rho_{\it{zz}}$) data using the equation,

 \begin{equation}
 \sigma_{\it{zx}} = -\frac{\rho_{\it{zx}}}{{\rho}^2_{\it{zz}} + {\rho}^2_{\it{zx}}}
 \label{Eq2}
 \end{equation}

Following the semiclassical two band model, we fitted the Hall conductivity  $\sigma_{\it{zx}}$ using the equation,

 \begin{equation}
 \centering
\sigma_{\it{zx}} = \left[ \frac{n_h{\mu}^2_h}{1 + (\mu_hB)^2} \:\:\text{--}\:\:\frac{n_e{\mu}^2_e}{1 + (\mu_eB)^2} \right] eB
\label{Eq3}
 \end{equation}

where $n_h (n_e)$ and $\mu_h (\mu_e)$ are hole (electron) density and mobility, respectively. The two-band fitting of $\sigma_{\it{zx}}$ at 2 K are shown in the top right inset of  Fig.~\ref{fig5}(c) and Fig.~\ref{fig5}(d) for HoFe$_6$Sn$_6$ and DyFe$_6$Sn$_6$, respectively. From  the Hall conductivity  fitting  of  HoFe$_6$Sn$_6$ at 2 K, the hole and electron density  are found to be $n_h=0.64\times 10^{19}$ $cm^{-3}$ and $n_e= 0.69\times 10^{19}$  $cm^{-3}$. The ratio of hole and electron density  $n_h/n_e= 0.93$, which is close to 1 for a perfect electron-hole compensation. Also, the hole and electron mobility are found to be $\mu_h=1.05\times 10^4$ $cm^2$ $V^{-1}$ $s^{-1}$ and $\mu_e = 3.62\times 10^4$ $cm^2$ $V^{-1}$ $s^{-1}$, respectively,  for  HoFe$_6$Sn$_6$. The mean mobility is estimated to be $\bar{\mu} = 1.5\times 10^4$ $cm^2$ $V^{-1}$ $s^{-1}$, using the relation $\bar{\mu} = \sqrt{{\mu_h}{\mu_e}}$ \cite{Ali2014} . The mean mobility in this system is  high  and comparable to the numerous other systems showing XMR ~\cite{Zheng2016, Han2017, Yu2017, Wang2018a, Niu2021}.

Further, in Fig.~\ref{fig5}(e)  we can notice that the mean carrier mobility rapidly decreases with increasing temperature between 2 and 100 K. Additionally, we also notice that the MR\% value drops between 2 and 100 K in a similar fashion of the mean mobility [see Fig.~\ref{fig5}(e)]. This observation suggests that the carrier mobility has a significant influence on the observed  XMR and ADMR [see Fig.~\ref{fig5_1}].  According to the earlier reports, XMR behaviour in  $\alpha-WP_2$  strongly depends on the high carrier mobility rather than electron-hole compensation, which is consistent with our observation in HoFe$_6$Sn$_6$ \cite{Lv2018}. Note that the band dispersions near the Fermi level controls the carrier mobility, and thus it is the electronic band structure playing the dominant role of ADMR in these systems. Further, DyFe$_6$Sn$_6$ also exhibits similar behaviour of carrier mobility to that of  HoFe$_6$Sn$_6$. In DyFe$_6$Sn$_6$ at 2 K, the hole and electron density  are found to be $n_h=3.95\times 10^{20}$ $cm^{-3}$ and $n_e= 3.77\times 10^{20}$  $cm^{-3}$, respectively. The ratio of hole and electron density is $n_h/n_e= 1.05$. The hole and electron mobility are found to be $\mu_h=0.10\times 10^4$ $cm^2$ $V^{-1}$ $s^{-1}$ and $\mu_e = 1.18\times 10^4$ $cm^2$ $V^{-1}$ $s^{-1}$, yielding the mean mobility of $\bar{\mu} = 0.35\times 10^4$ $cm^2$ $V^{-1}$ $s^{-1}$.

\section{Summary}

In conclusion, we systematically studied the magnetic and magnetotransport properties on the high-quality single crystals of the Fe-based RFe$_6$Sn$_6$ (R = Ho, Dy) kagome systems. Extremely large magnetoresistance (XMR) is observed at low temperatures reaching the maximum MR percentage as high as $3\times 10^{3} $ $\%$ for HoFe$_6$Sn$_6$ and  $1\times 10^{3}$ $\%$ for DyFe$_6$Sn$_6$ at 2 K under 9 T of applied magnetic field.  Hall effect measurements demonstrate the electron-hole charge compensation and high-carrier mobility, leading to extremely large magnetoresistance at low temperatures. Further, the angle-dependent magnetoresistance (ADMR) data reveal high anisotropy in the magnetoresistance. Importantly, the magnetoresistance anisotropic pattern changes significantly with temperature, implies the temperature dependence of the Fermi surface topology in these systems.

\section{Acknowledgement}

N.D. acknowledges IoE-IISc Postdoctoral Fellowship. S.G. acknowledges the University Grants Commission (UGC), India for the Ph.D. fellowship.  A.N. acknowledges support from DST Nano Mission (project no. DST/NM/TUE/QM-10/2019(G)/1). S.T. thanks  Anusandhan National Research Foundation (ANRF), India, through Grant no. CRG/2023/000748. S.T. thanks SERB (DST), India for the financial support through Grant no. SRG/2020/00393. The experimental work conducted in this research utilized the Technical Research Centre (TRC) Instrument Facilities at S. N. Bose National Centre for Basic Sciences, established as part of the TRC project funded by the Department of Science and Technology, Government of India.

\bibliographystyle{model1-num-names}
\bibliography{RFe6Sn6_clean.bib}

\begin{thebibliography}{74}
\expandafter\ifx\csname natexlab\endcsname\relax\def\natexlab#1{#1}\fi
\providecommand{\url}[1]{\texttt{#1}}
\providecommand{\href}[2]{#2}
\providecommand{\path}[1]{#1}
\providecommand{\DOIprefix}{doi:}
\providecommand{\ArXivprefix}{arXiv:}
\providecommand{\URLprefix}{URL: }
\providecommand{\Pubmedprefix}{pmid:}
\providecommand{\doi}[1]{\href{http://dx.doi.org/#1}{\path{#1}}}
\providecommand{\Pubmed}[1]{\href{pmid:#1}{\path{#1}}}
\providecommand{\bibinfo}[2]{#2}
\ifx\xfnm\relax \def\xfnm[#1]{\unskip,\space#1}\fi
\bibitem[{Heaps(1939)}]{Heaps1939}
\bibinfo{author}{C.~W. Heaps},
\newblock \bibinfo{title}{{The Magnetoresistance of Nickel in Large Fields}},
\newblock \bibinfo{journal}{Physical Review} \bibinfo{volume}{55}
  (\bibinfo{year}{1939}) \bibinfo{pages}{1069--1071}.
\bibitem[{Parkin et~al.(1990)Parkin, More, and Roche}]{Parkin1990}
\bibinfo{author}{S.~S.~P. Parkin}, \bibinfo{author}{N.~More},
  \bibinfo{author}{K.~P. Roche},
\newblock \bibinfo{title}{{Oscillations in exchange coupling and
  magnetoresistance in metallic superlattice structures: Co/Ru, Co/Cr, and
  Fe/Cr}},
\newblock \bibinfo{journal}{Phys. Rev. Lett.} \bibinfo{volume}{64}
  (\bibinfo{year}{1990}) \bibinfo{pages}{2304--2307}.
\bibitem[{Niu and Zhu(2021)}]{Niu2021}
\bibinfo{author}{R.~Niu}, \bibinfo{author}{W.~Zhu},
\newblock \bibinfo{title}{{Materials and possible mechanisms of extremely large
  magnetoresistance: a review}},
\newblock \bibinfo{journal}{J. Phys. Condens. Matter} \bibinfo{volume}{34}
  (\bibinfo{year}{2021}) \bibinfo{pages}{113001}.
\bibitem[{Ramirez(1997)}]{Ramirez1997}
\bibinfo{author}{A.~P. Ramirez},
\newblock \bibinfo{title}{{Colossal magnetoresistance}},
\newblock \bibinfo{journal}{J. Phys. Condens. Matter} \bibinfo{volume}{9}
  (\bibinfo{year}{1997}) \bibinfo{pages}{8171--8199}.
\bibitem[{Yang and Zhang(2021)}]{Yang2021}
\bibinfo{author}{S.~Yang}, \bibinfo{author}{J.~Zhang},
\newblock \bibinfo{title}{{Current progress of magnetoresistance sensors}},
\newblock \bibinfo{journal}{Chemosensors} \bibinfo{volume}{9}
  (\bibinfo{year}{2021}) \bibinfo{pages}{211}.
\bibitem[{Vitayaya et~al.(2024)Vitayaya, Nehan, Munazat, Manawan, and
  Kurniawan}]{Vitayaya2024}
\bibinfo{author}{O.~Vitayaya}, \bibinfo{author}{P.~Z.~Z. Nehan},
  \bibinfo{author}{D.~R. Munazat}, \bibinfo{author}{M.~T.~E. Manawan},
  \bibinfo{author}{B.~Kurniawan},
\newblock \bibinfo{title}{{Magnetoresistance (MR) properties of magnetic
  materials}},
\newblock \bibinfo{journal}{RSC Advances} \bibinfo{volume}{14}
  (\bibinfo{year}{2024}) \bibinfo{pages}{18617--18645}.
\bibitem[{Pippard(1989)}]{Pippard1989}
\bibinfo{author}{A.~B. Pippard}, \bibinfo{title}{Magnetoresistance in metals},
  volume~\bibinfo{volume}{2}, \bibinfo{publisher}{Cambridge university press},
  \bibinfo{year}{1989}.
\bibitem[{Blundell(2001)}]{Blundell2001}
\bibinfo{author}{S.~Blundell}, \bibinfo{title}{{Magnetism in condensed
  matter}}, \bibinfo{publisher}{OUP Oxford}, \bibinfo{year}{2001}.
\bibitem[{Nickel(1995)}]{Nickel1995}
\bibinfo{author}{J.~Nickel}, \bibinfo{title}{Magnetoresistance overview},
  \bibinfo{publisher}{Hewlett-Packard Laboratories, Technical Publications
  Department Palo Alto~…}, \bibinfo{year}{1995}.
\bibitem[{Tsymbal and Pettifor(2001)}]{Tsymbal2001}
\bibinfo{author}{E.~Tsymbal}, \bibinfo{author}{D.~Pettifor},
  \bibinfo{title}{{Perspectives of giant magnetoresistance}},
  \bibinfo{year}{2001}. \URLprefix
  \url{https://doi.org/10.1016/s0081-1947(01)80019-9}.
  \DOIprefix\doi{10.1016/s0081-1947(01)80019-9}.
\bibitem[{Inoue and Maekawa(1996)}]{Inoue1996}
\bibinfo{author}{J.~Inoue}, \bibinfo{author}{S.~Maekawa},
\newblock \bibinfo{title}{{Theory of tunneling magnetoresistance in granular
  magnetic films}},
\newblock \bibinfo{journal}{Phys. Rev. B} \bibinfo{volume}{53}
  (\bibinfo{year}{1996}) \bibinfo{pages}{R11927--R11929}.
\bibitem[{Ali et~al.(2014)Ali, Xiong, Flynn, Tao, Gibson, Schoop, Liang,
  Haldolaarachchige, Hirschberger, Ong, and Cava}]{Ali2014}
\bibinfo{author}{M.~N. Ali}, \bibinfo{author}{J.~Xiong},
  \bibinfo{author}{S.~Flynn}, \bibinfo{author}{J.~Tao},
  \bibinfo{author}{Q.~Gibson}, \bibinfo{author}{L.~M. Schoop},
  \bibinfo{author}{T.~Liang}, \bibinfo{author}{N.~Haldolaarachchige},
  \bibinfo{author}{M.~Hirschberger}, \bibinfo{author}{N.~P. Ong},
  \bibinfo{author}{R.~J. Cava},
\newblock \bibinfo{title}{{Large, non-saturating magnetoresistance in
  $WTe_2$}},
\newblock \bibinfo{journal}{Nature} \bibinfo{volume}{514}
  (\bibinfo{year}{2014}) \bibinfo{pages}{205--208}.
\bibitem[{Han et~al.(2017)Han, Xu, Botana, Xiao, Wang, Yang, Chung, Kanatzidis,
  Norman, Crabtree, and Kwok}]{Han2017}
\bibinfo{author}{F.~Han}, \bibinfo{author}{J.~Xu}, \bibinfo{author}{A.~S.
  Botana}, \bibinfo{author}{Z.~L. Xiao}, \bibinfo{author}{Y.~L. Wang},
  \bibinfo{author}{W.~G. Yang}, \bibinfo{author}{D.~Y. Chung},
  \bibinfo{author}{M.~G. Kanatzidis}, \bibinfo{author}{M.~R. Norman},
  \bibinfo{author}{G.~W. Crabtree}, \bibinfo{author}{W.~K. Kwok},
\newblock \bibinfo{title}{{Separation of electron and hole dynamics in the
  semimetal LaSb}},
\newblock \bibinfo{journal}{Phys. Rev. B} \bibinfo{volume}{96}
  (\bibinfo{year}{2017}) \bibinfo{pages}{125112}.
\bibitem[{Gao et~al.(2017)Gao, Hao, Zheng, Ning, Wu, Zhu, Zheng, Zhang, Lu,
  Zhang, Xi, Yang, Du, Zhang, Zhang, and Tian}]{Gao2017}
\bibinfo{author}{W.~Gao}, \bibinfo{author}{N.~Hao}, \bibinfo{author}{F.-W.
  Zheng}, \bibinfo{author}{W.~Ning}, \bibinfo{author}{M.~Wu},
  \bibinfo{author}{X.~Zhu}, \bibinfo{author}{G.~Zheng},
  \bibinfo{author}{J.~Zhang}, \bibinfo{author}{J.~Lu},
  \bibinfo{author}{H.~Zhang}, \bibinfo{author}{C.~Xi},
  \bibinfo{author}{J.~Yang}, \bibinfo{author}{H.~Du},
  \bibinfo{author}{P.~Zhang}, \bibinfo{author}{Y.~Zhang},
  \bibinfo{author}{M.~Tian},
\newblock \bibinfo{title}{{Extremely Large Magnetoresistance in a Topological
  Semimetal Candidate Pyrite ${\mathrm{PtBi}}_{2}$}},
\newblock \bibinfo{journal}{Phys. Rev. Lett.} \bibinfo{volume}{118}
  (\bibinfo{year}{2017}) \bibinfo{pages}{256601}.
\bibitem[{Okawa et~al.(2018)Okawa, Kanou, Namiki, and Sasagawa}]{Okawa2018}
\bibinfo{author}{K.~Okawa}, \bibinfo{author}{M.~Kanou},
  \bibinfo{author}{H.~Namiki}, \bibinfo{author}{T.~Sasagawa},
\newblock \bibinfo{title}{{Extremely large magnetoresistance induced by hidden
  three-dimensional Dirac bands in nonmagnetic semimetal InBi}},
\newblock \bibinfo{journal}{Phys. Rev. Mater.} \bibinfo{volume}{2}
  (\bibinfo{year}{2018}) \bibinfo{pages}{124201}.
\bibitem[{Shekhar et~al.(2015)Shekhar, Nayak, Sun, Schmidt, Nicklas,
  Leermakers, Zeitler, Skourski, Wosnitza, Liu, Chen, Schnelle, Borrmann, Grin,
  Felser, and Yan}]{Shekhar2015}
\bibinfo{author}{C.~Shekhar}, \bibinfo{author}{A.~K. Nayak},
  \bibinfo{author}{Y.~Sun}, \bibinfo{author}{M.~Schmidt},
  \bibinfo{author}{M.~Nicklas}, \bibinfo{author}{I.~Leermakers},
  \bibinfo{author}{U.~Zeitler}, \bibinfo{author}{Y.~Skourski},
  \bibinfo{author}{J.~Wosnitza}, \bibinfo{author}{Z.~Liu},
  \bibinfo{author}{Y.~Chen}, \bibinfo{author}{W.~Schnelle},
  \bibinfo{author}{H.~Borrmann}, \bibinfo{author}{Y.~Grin},
  \bibinfo{author}{C.~Felser}, \bibinfo{author}{B.~Yan},
\newblock \bibinfo{title}{{Extremely large magnetoresistance and ultrahigh
  mobility in the topological Weyl semimetal candidate NbP}},
\newblock \bibinfo{journal}{Nat. Phys.} \bibinfo{volume}{11}
  (\bibinfo{year}{2015}) \bibinfo{pages}{645--649}.
\bibitem[{Mondal et~al.(2020)Mondal, Sasmal, Kulkarni, Maurya, Nakamura, Aoki,
  Harima, and Thamizhavel}]{Mondal2020}
\bibinfo{author}{R.~Mondal}, \bibinfo{author}{S.~Sasmal},
  \bibinfo{author}{R.~Kulkarni}, \bibinfo{author}{A.~Maurya},
  \bibinfo{author}{A.~Nakamura}, \bibinfo{author}{D.~Aoki},
  \bibinfo{author}{H.~Harima}, \bibinfo{author}{A.~Thamizhavel},
\newblock \bibinfo{title}{{Extremely large magnetoresistance, anisotropic Hall
  effect, and Fermi surface topology in single-crystalline
  $\mathrm{W}{\mathrm{Si}}_{2}$}},
\newblock \bibinfo{journal}{Phys. Rev. B} \bibinfo{volume}{102}
  (\bibinfo{year}{2020}) \bibinfo{pages}{115158}.
\bibitem[{Zhang et~al.(2024)Zhang, Li, Chen, Li, and Uher}]{Zhang2024}
\bibinfo{author}{Y.~Zhang}, \bibinfo{author}{Z.~Li}, \bibinfo{author}{K.-W.
  Chen}, \bibinfo{author}{L.~Li}, \bibinfo{author}{C.~Uher},
\newblock \bibinfo{title}{{Extremely large magnetoresistance and quantum
  oscillations in semimetal $Ni_3In_2S_2$}},
\newblock \bibinfo{journal}{Mater. Today Phys.} \bibinfo{volume}{40}
  (\bibinfo{year}{2024}) \bibinfo{pages}{101318}.
\bibitem[{Morali et~al.(2019)Morali, Batabyal, Nag, Liu, Xu, Sun, Yan, Felser,
  Avraham, and Beidenkopf}]{Morali2019}
\bibinfo{author}{N.~Morali}, \bibinfo{author}{R.~Batabyal},
  \bibinfo{author}{P.~K. Nag}, \bibinfo{author}{E.~Liu},
  \bibinfo{author}{Q.~Xu}, \bibinfo{author}{Y.~Sun}, \bibinfo{author}{B.~Yan},
  \bibinfo{author}{C.~Felser}, \bibinfo{author}{N.~Avraham},
  \bibinfo{author}{H.~Beidenkopf},
\newblock \bibinfo{title}{{Fermi-arc diversity on surface terminations of the
  magnetic Weyl semimetal Co$_3$Sn$_2$S$_2$}},
\newblock \bibinfo{journal}{Science} \bibinfo{volume}{365}
  (\bibinfo{year}{2019}) \bibinfo{pages}{1286--1291}.
\bibitem[{Wang et~al.(2018)Wang, Xu, Lou, Liu, Li, Huang, Shen, Weng, Wang, and
  Lei}]{Wang2018}
\bibinfo{author}{Q.~Wang}, \bibinfo{author}{Y.~Xu}, \bibinfo{author}{R.~Lou},
  \bibinfo{author}{Z.~Liu}, \bibinfo{author}{M.~Li},
  \bibinfo{author}{Y.~Huang}, \bibinfo{author}{D.~Shen},
  \bibinfo{author}{H.~Weng}, \bibinfo{author}{S.~Wang},
  \bibinfo{author}{H.~Lei},
\newblock \bibinfo{title}{{Large intrinsic anomalous Hall effect in
  half-metallic ferromagnet $Co_3Sn_2S_2$ with magnetic Weyl fermions}},
\newblock \bibinfo{journal}{Nat. Commun.} \bibinfo{volume}{9}
  (\bibinfo{year}{2018}).
\bibitem[{Nakatsuji et~al.(2015)Nakatsuji, Kiyohara, and Higo}]{Nakatsuji2015}
\bibinfo{author}{S.~Nakatsuji}, \bibinfo{author}{N.~Kiyohara},
  \bibinfo{author}{T.~Higo},
\newblock \bibinfo{title}{{Large anomalous Hall effect in a non-collinear
  antiferromagnet at room temperature}},
\newblock \bibinfo{journal}{Nature} \bibinfo{volume}{527}
  (\bibinfo{year}{2015}) \bibinfo{pages}{212--215}.
\bibitem[{Nayak et~al.(2016)Nayak, Fischer, Sun, Yan, Karel, Komarek, Shekhar,
  Kumar, Schnelle, K{\"u}bler et~al.}]{Nayak2016}
\bibinfo{author}{A.~K. Nayak}, \bibinfo{author}{J.~E. Fischer},
  \bibinfo{author}{Y.~Sun}, \bibinfo{author}{B.~Yan},
  \bibinfo{author}{J.~Karel}, \bibinfo{author}{A.~C. Komarek},
  \bibinfo{author}{C.~Shekhar}, \bibinfo{author}{N.~Kumar},
  \bibinfo{author}{W.~Schnelle}, \bibinfo{author}{J.~K{\"u}bler}, et~al.,
\newblock \bibinfo{title}{{Large anomalous Hall effect driven by a nonvanishing
  Berry curvature in the noncolinear antiferromagnet Mn$_3$Ge}},
\newblock \bibinfo{journal}{Sci. Adv.} \bibinfo{volume}{2}
  (\bibinfo{year}{2016}) \bibinfo{pages}{e1501870}.
\bibitem[{K{\"u}bler and Felser(2018)}]{Kuebler2018}
\bibinfo{author}{J.~K{\"u}bler}, \bibinfo{author}{C.~Felser},
\newblock \bibinfo{title}{{Weyl fermions in antiferromagnetic $Mn_3Sn$ and
  $Mn_3Ge$}},
\newblock \bibinfo{journal}{EPL (Europhysics Letters)} \bibinfo{volume}{120}
  (\bibinfo{year}{2018}) \bibinfo{pages}{47002}.
\bibitem[{Low et~al.(2022)Low, Ghosh, Changdar, Routh, Purwar, and
  Thirupathaiah}]{Low2022}
\bibinfo{author}{A.~Low}, \bibinfo{author}{S.~Ghosh},
  \bibinfo{author}{S.~Changdar}, \bibinfo{author}{S.~Routh},
  \bibinfo{author}{S.~Purwar}, \bibinfo{author}{S.~Thirupathaiah},
\newblock \bibinfo{title}{{Tuning of topological properties in the strongly
  correlated antiferromagnet $Mn_3Sn$ via Fe doping}},
\newblock \bibinfo{journal}{Phys. Rev. B} \bibinfo{volume}{106}
  (\bibinfo{year}{2022}) \bibinfo{pages}{144429}.
\bibitem[{Changdar et~al.(2023)Changdar, Ghosh, Bose, Kar, Low, Fèvre,
  Bertran, Narayan, and Thirupathaiah}]{Changdar2023}
\bibinfo{author}{S.~Changdar}, \bibinfo{author}{S.~Ghosh},
  \bibinfo{author}{A.~Bose}, \bibinfo{author}{I.~Kar},
  \bibinfo{author}{A.~Low}, \bibinfo{author}{P.~L. Fèvre},
  \bibinfo{author}{F.~Bertran}, \bibinfo{author}{A.~Narayan},
  \bibinfo{author}{S.~Thirupathaiah},
\newblock \bibinfo{title}{{Weak electronic correlations observed in magnetic
  Weyl Semimetal Mn$_3$Ge}},
\newblock \bibinfo{journal}{J. Phys. Condens. Matter} \bibinfo{volume}{36}
  (\bibinfo{year}{2023}) \bibinfo{pages}{125502}.
\bibitem[{Ghosh et~al.(2023)Ghosh, Low, Ghorai, Mandal, and
  Thirupathaiah}]{Ghosh2023}
\bibinfo{author}{S.~Ghosh}, \bibinfo{author}{A.~Low},
  \bibinfo{author}{S.~Ghorai}, \bibinfo{author}{K.~Mandal},
  \bibinfo{author}{S.~Thirupathaiah},
\newblock \bibinfo{title}{{Tuning of electrical, magnetic, and topological
  properties of magnetic Weyl semimetal $Mn_{3+x}Ge$ by Fe doping}},
\newblock \bibinfo{journal}{J. Phys. Condens. Matter} \bibinfo{volume}{35}
  (\bibinfo{year}{2023}) \bibinfo{pages}{485701}.
\bibitem[{Jiang et~al.(2021)Jiang, Yin, Denner, Shumiya, Ortiz, Xu, Guguchia,
  He, Hossain, Liu, Ruff, Kautzsch, Zhang, Chang, Belopolski, Zhang, Cochran,
  Multer, Litskevich, Cheng, Yang, Wang, Thomale, Neupert, Wilson, and
  Hasan}]{Jiang2021}
\bibinfo{author}{Y.-X. Jiang}, \bibinfo{author}{J.-X. Yin},
  \bibinfo{author}{M.~M. Denner}, \bibinfo{author}{N.~Shumiya},
  \bibinfo{author}{B.~R. Ortiz}, \bibinfo{author}{G.~Xu},
  \bibinfo{author}{Z.~Guguchia}, \bibinfo{author}{J.~He},
  \bibinfo{author}{S.~Hossain}, \bibinfo{author}{X.~Liu},
  \bibinfo{author}{J.~Ruff}, \bibinfo{author}{L.~Kautzsch},
  \bibinfo{author}{S.~S. Zhang}, \bibinfo{author}{G.~Chang},
  \bibinfo{author}{I.~Belopolski}, \bibinfo{author}{Q.~Zhang},
  \bibinfo{author}{T.~A. Cochran}, \bibinfo{author}{D.~Multer},
  \bibinfo{author}{M.~Litskevich}, \bibinfo{author}{Z.-J. Cheng},
  \bibinfo{author}{X.~P. Yang}, \bibinfo{author}{Z.~Wang},
  \bibinfo{author}{R.~Thomale}, \bibinfo{author}{T.~Neupert},
  \bibinfo{author}{S.~D. Wilson}, \bibinfo{author}{M.~Z. Hasan},
\newblock \bibinfo{title}{{Unconventional chiral charge order in kagome
  superconductor $KV_3Sb_5$}},
\newblock \bibinfo{journal}{Nat. Mater.} \bibinfo{volume}{20}
  (\bibinfo{year}{2021}) \bibinfo{pages}{1353--1357}.
\bibitem[{Hu et~al.(2022)Hu, Wu, Ortiz, Ju, Han, Z, Plumb, Radovic, Thomale,
  Wilson, Schnyder, and Shi}]{Hu2022}
\bibinfo{author}{Y.~Hu}, \bibinfo{author}{X.~Wu}, \bibinfo{author}{B.~R.
  Ortiz}, \bibinfo{author}{S.~Ju}, \bibinfo{author}{X.~Han},
  \bibinfo{author}{J.~Z, MA}, \bibinfo{author}{N.~C. Plumb},
  \bibinfo{author}{M.~Radovic}, \bibinfo{author}{R.~Thomale},
  \bibinfo{author}{S.~D. Wilson}, \bibinfo{author}{A.~P. Schnyder},
  \bibinfo{author}{M.~Shi},
\newblock \bibinfo{title}{{Rich nature of Van Hove singularities in Kagome
  superconductor $CsV_3Sb_5$}},
\newblock \bibinfo{journal}{Nat. Commun.} \bibinfo{volume}{13}
  (\bibinfo{year}{2022}).
\bibitem[{Ma et~al.(2021)Ma, Xu, Yin, Yang, Zhou, Cheng, Huang, Qu, Wang,
  Hasan, and Jia}]{Ma2021}
\bibinfo{author}{W.~Ma}, \bibinfo{author}{X.~Xu}, \bibinfo{author}{J.-X. Yin},
  \bibinfo{author}{H.~Yang}, \bibinfo{author}{H.~Zhou}, \bibinfo{author}{Z.-J.
  Cheng}, \bibinfo{author}{Y.~Huang}, \bibinfo{author}{Z.~Qu},
  \bibinfo{author}{F.~Wang}, \bibinfo{author}{M.~Z. Hasan},
  \bibinfo{author}{S.~Jia},
\newblock \bibinfo{title}{{Rare Earth Engineering in
  $R{\mathrm{Mn}}_{6}{\mathrm{Sn}}_{6}$
  ($R=\text{Gd}\text{\ensuremath{-}}\text{Tm}$, Lu) Topological Kagome
  Magnets}},
\newblock \bibinfo{journal}{Phys. Rev. Lett.} \bibinfo{volume}{126}
  (\bibinfo{year}{2021}) \bibinfo{pages}{246602}.
\bibitem[{Dhakal et~al.(2021)Dhakal, Cheenicode~Kabeer, Pathak, Kabir, Poudel,
  Filippone, Casey, Pradhan~Sakhya, Regmi, Sims, Dimitri, Manfrinetti, Gofryk,
  Oppeneer, and Neupane}]{Dhakal2021}
\bibinfo{author}{G.~Dhakal}, \bibinfo{author}{F.~Cheenicode~Kabeer},
  \bibinfo{author}{A.~K. Pathak}, \bibinfo{author}{F.~Kabir},
  \bibinfo{author}{N.~Poudel}, \bibinfo{author}{R.~Filippone},
  \bibinfo{author}{J.~Casey}, \bibinfo{author}{A.~Pradhan~Sakhya},
  \bibinfo{author}{S.~Regmi}, \bibinfo{author}{C.~Sims},
  \bibinfo{author}{K.~Dimitri}, \bibinfo{author}{P.~Manfrinetti},
  \bibinfo{author}{K.~Gofryk}, \bibinfo{author}{P.~M. Oppeneer},
  \bibinfo{author}{M.~Neupane},
\newblock \bibinfo{title}{{Anisotropically large anomalous and topological Hall
  effect in a kagome magnet}},
\newblock \bibinfo{journal}{Phys. Rev. B} \bibinfo{volume}{104}
  (\bibinfo{year}{2021}) \bibinfo{pages}{L161115}.
\bibitem[{Gao et~al.(2021)Gao, Shen, Wang, Shi, Zhao, Li, Cao, Pei, Ge, Li, Li,
  Chen, Yan, and Qi}]{Gao2021}
\bibinfo{author}{L.~Gao}, \bibinfo{author}{S.~Shen}, \bibinfo{author}{Q.~Wang},
  \bibinfo{author}{W.~Shi}, \bibinfo{author}{Y.~Zhao}, \bibinfo{author}{C.~Li},
  \bibinfo{author}{W.~Cao}, \bibinfo{author}{C.~Pei}, \bibinfo{author}{J.-Y.
  Ge}, \bibinfo{author}{G.~Li}, \bibinfo{author}{J.~Li},
  \bibinfo{author}{Y.~Chen}, \bibinfo{author}{S.~Yan}, \bibinfo{author}{Y.~Qi},
\newblock \bibinfo{title}{{Anomalous Hall effect in ferrimagnetic metal
  $RMn_6Sn_6$ (R = Tb, Dy, Ho) with clean Mn kagome lattice}},
\newblock \bibinfo{journal}{Appl. Phys. Lett.} \bibinfo{volume}{119}
  (\bibinfo{year}{2021}).
\bibitem[{Zhou et~al.(2023)Zhou, Shi, Huang, Ma, Xu, Wang, and Jia}]{Zhou2023}
\bibinfo{author}{H.~Zhou}, \bibinfo{author}{M.~Shi},
  \bibinfo{author}{Y.~Huang}, \bibinfo{author}{W.~Ma}, \bibinfo{author}{X.~Xu},
  \bibinfo{author}{J.~Wang}, \bibinfo{author}{S.~Jia},
\newblock \bibinfo{title}{{Metamagnetic transition and anomalous Hall effect in
  Mn-based kagom\'e magnets
  $R{\mathrm{Mn}}_{6}{\mathrm{Ge}}_{6}(R=\mathrm{Tb}\text{\ensuremath{-}}\mathrm{Lu})$}},
\newblock \bibinfo{journal}{Phys. Rev. Mater.} \bibinfo{volume}{7}
  (\bibinfo{year}{2023}) \bibinfo{pages}{024404}.
\bibitem[{Xu et~al.(2021)Xu, Heitmann, Zhang, Xu, and Ke}]{Xu2021}
\bibinfo{author}{C.~Q. Xu}, \bibinfo{author}{T.~W. Heitmann},
  \bibinfo{author}{H.~Zhang}, \bibinfo{author}{X.~Xu}, \bibinfo{author}{X.~Ke},
\newblock \bibinfo{title}{{Magnetic phase transition, magnetoresistance, and
  anomalous Hall effect in Ga-substituted
  $\mathrm{Y}{\mathrm{Mn}}_{6}{\mathrm{Sn}}_{6}$ with a ferromagnetic kagome
  lattice}},
\newblock \bibinfo{journal}{Phys. Rev. B} \bibinfo{volume}{104}
  (\bibinfo{year}{2021}) \bibinfo{pages}{024413}.
\bibitem[{Kabir et~al.(2022)Kabir, Filippone, Dhakal, Lee, Poudel, Casey,
  Sakhya, Regmi, Smith, Manfrinetti, Ke, Gofryk, Neupane, and
  Pathak}]{Kabir2022}
\bibinfo{author}{F.~Kabir}, \bibinfo{author}{R.~Filippone},
  \bibinfo{author}{G.~Dhakal}, \bibinfo{author}{Y.~Lee},
  \bibinfo{author}{N.~Poudel}, \bibinfo{author}{J.~Casey},
  \bibinfo{author}{A.~P. Sakhya}, \bibinfo{author}{S.~Regmi},
  \bibinfo{author}{R.~Smith}, \bibinfo{author}{P.~Manfrinetti},
  \bibinfo{author}{L.~Ke}, \bibinfo{author}{K.~Gofryk},
  \bibinfo{author}{M.~Neupane}, \bibinfo{author}{A.~K. Pathak},
\newblock \bibinfo{title}{Unusual magnetic and transport properties in
  ${\mathrm{homn}}_{6}{\mathrm{sn}}_{6}$ kagome magnet},
\newblock \bibinfo{journal}{Phys. Rev. Mater.} \bibinfo{volume}{6}
  (\bibinfo{year}{2022}) \bibinfo{pages}{064404}.
\bibitem[{Liu et~al.(2023)Liu, Zhang, Li, Yan, Zhang, Hou, and Fu}]{Liu2023a}
\bibinfo{author}{C.~Liu}, \bibinfo{author}{H.~Zhang}, \bibinfo{author}{Z.~Li},
  \bibinfo{author}{Y.~Yan}, \bibinfo{author}{Y.~Zhang},
  \bibinfo{author}{Z.~Hou}, \bibinfo{author}{X.~Fu},
\newblock \bibinfo{title}{{Nontrivial spin textures induced remarkable
  topological Hall effect and extraordinary magnetoresistance in kagome magnet
  $TmMn_6Sn_6$}},
\newblock \bibinfo{journal}{Surfaces and Interfaces} \bibinfo{volume}{39}
  (\bibinfo{year}{2023}) \bibinfo{pages}{102866}.
\bibitem[{Low et~al.(2024)Low, Bhowmik, Ghosh, and Thirupathaiah}]{Low2024}
\bibinfo{author}{A.~Low}, \bibinfo{author}{T.~K. Bhowmik},
  \bibinfo{author}{S.~Ghosh}, \bibinfo{author}{S.~Thirupathaiah},
\newblock \bibinfo{title}{{Anisotropic nonsaturating magnetoresistance observed
  in ${\mathrm{HoMn}}_{6}{\mathrm{Ge}}_{6}$: A kagome Dirac semimetal}},
\newblock \bibinfo{journal}{Phys. Rev. B} \bibinfo{volume}{109}
  (\bibinfo{year}{2024}) \bibinfo{pages}{195104}.
\bibitem[{Peng et~al.(2021)Peng, Han, Pokharel, Shen, Li, Hashimoto, Lu, Ortiz,
  Luo, Li, Guo, Wang, Cui, Sun, Qiao, Wilson, and He}]{Peng2021}
\bibinfo{author}{S.~Peng}, \bibinfo{author}{Y.~Han},
  \bibinfo{author}{G.~Pokharel}, \bibinfo{author}{J.~Shen},
  \bibinfo{author}{Z.~Li}, \bibinfo{author}{M.~Hashimoto},
  \bibinfo{author}{D.~Lu}, \bibinfo{author}{B.~R. Ortiz},
  \bibinfo{author}{Y.~Luo}, \bibinfo{author}{H.~Li}, \bibinfo{author}{M.~Guo},
  \bibinfo{author}{B.~Wang}, \bibinfo{author}{S.~Cui},
  \bibinfo{author}{Z.~Sun}, \bibinfo{author}{Z.~Qiao}, \bibinfo{author}{S.~D.
  Wilson}, \bibinfo{author}{J.~He},
\newblock \bibinfo{title}{{Realizing Kagome Band Structure in Two-Dimensional
  Kagome Surface States of $R{\mathrm{V}}_{6}{\mathrm{Sn}}_{6}$
  ($R=\mathrm{Gd}$, Ho)}},
\newblock \bibinfo{journal}{Phys. Rev. Lett.} \bibinfo{volume}{127}
  (\bibinfo{year}{2021}) \bibinfo{pages}{266401}.
\bibitem[{Zhang et~al.(2022)Zhang, Liu, Cui, Guo, Wang, Shi, Zhang, Wang, Dong,
  Sun, Dun, and Cheng}]{Zhang2022}
\bibinfo{author}{X.~Zhang}, \bibinfo{author}{Z.~Liu}, \bibinfo{author}{Q.~Cui},
  \bibinfo{author}{Q.~Guo}, \bibinfo{author}{N.~Wang},
  \bibinfo{author}{L.~Shi}, \bibinfo{author}{H.~Zhang},
  \bibinfo{author}{W.~Wang}, \bibinfo{author}{X.~Dong},
  \bibinfo{author}{J.~Sun}, \bibinfo{author}{Z.~Dun},
  \bibinfo{author}{J.~Cheng},
\newblock \bibinfo{title}{{Electronic and magnetic properties of intermetallic
  kagome magnets
  $R{\mathrm{V}}_{6}{\mathrm{Sn}}_{6}(R=\mathrm{Tb}\text{\ensuremath{-}}\mathrm{Tm})$}},
\newblock \bibinfo{journal}{Phys. Rev. Mater.} \bibinfo{volume}{6}
  (\bibinfo{year}{2022}) \bibinfo{pages}{105001}.
\bibitem[{Lee and Mun(2022)}]{Lee2022}
\bibinfo{author}{J.~Lee}, \bibinfo{author}{E.~Mun},
\newblock \bibinfo{title}{{Anisotropic magnetic property of single crystals
  $R{\mathrm{V}}_{6}{\mathrm{Sn}}_{6}$ $(R=\mathrm{Y},
  \mathrm{Gd}\text{\ensuremath{-}}\mathrm{Tm}, \mathrm{Lu})$}},
\newblock \bibinfo{journal}{Phys. Rev. Mater.} \bibinfo{volume}{6}
  (\bibinfo{year}{2022}) \bibinfo{pages}{083401}.
\bibitem[{Arachchige et~al.(2022)Arachchige, Meier, Marshall, Matsuoka, Xue,
  McGuire, Hermann, Cao, and Mandrus}]{Arachchige2022}
\bibinfo{author}{H.~W.~S. Arachchige}, \bibinfo{author}{W.~R. Meier},
  \bibinfo{author}{M.~Marshall}, \bibinfo{author}{T.~Matsuoka},
  \bibinfo{author}{R.~Xue}, \bibinfo{author}{M.~A. McGuire},
  \bibinfo{author}{R.~P. Hermann}, \bibinfo{author}{H.~Cao},
  \bibinfo{author}{D.~Mandrus},
\newblock \bibinfo{title}{{Charge Density Wave in Kagome Lattice Intermetallic
  ${\mathrm{ScV}}_{6}{\mathrm{Sn}}_{6}$}},
\newblock \bibinfo{journal}{Phys. Rev. Lett.} \bibinfo{volume}{129}
  (\bibinfo{year}{2022}) \bibinfo{pages}{216402}.
\bibitem[{Liu et~al.(2023)Liu, Lyu, Liu, Zhang, Yang, Du, Wang, Wei, and
  Liu}]{Liu2023}
\bibinfo{author}{Y.~Liu}, \bibinfo{author}{M.~Lyu}, \bibinfo{author}{J.~Liu},
  \bibinfo{author}{S.~Zhang}, \bibinfo{author}{J.~Yang},
  \bibinfo{author}{Z.~Du}, \bibinfo{author}{B.~Wang}, \bibinfo{author}{H.~Wei},
  \bibinfo{author}{E.~Liu},
\newblock \bibinfo{title}{{Structural Determination, Unstable
  Antiferromagnetism and Transport Properties of Fe-Kagome $Y_{0. 5}Fe_3Sn_3$
  Single Crystals}},
\newblock \bibinfo{journal}{Chinese Phys. Lett.} \bibinfo{volume}{40}
  (\bibinfo{year}{2023}) \bibinfo{pages}{047102}.
\bibitem[{Giannozzi et~al.(2009)Giannozzi, Baroni, Bonini, Calandra, Car,
  Cavazzoni, Ceresoli, Chiarotti, Cococcioni, Dabo, Corso, De~Gironcoli,
  Fabris, Fratesi, Gebauer, Gerstmann, Gougoussis, Kokalj, Lazzeri,
  Martin-Samos, Marzari, Mauri, Mazzarello, Paolini, Pasquarello, Paulatto,
  Sbraccia, Scandolo, Sclauzero, Seitsonen, Smogunov, Umari, and
  Wentzcovitch}]{Giannozzi2009}
\bibinfo{author}{P.~Giannozzi}, \bibinfo{author}{S.~Baroni},
  \bibinfo{author}{N.~Bonini}, \bibinfo{author}{M.~Calandra},
  \bibinfo{author}{R.~Car}, \bibinfo{author}{C.~Cavazzoni},
  \bibinfo{author}{D.~Ceresoli}, \bibinfo{author}{G.~L. Chiarotti},
  \bibinfo{author}{M.~Cococcioni}, \bibinfo{author}{I.~Dabo},
  \bibinfo{author}{A.~D. Corso}, \bibinfo{author}{S.~De~Gironcoli},
  \bibinfo{author}{S.~Fabris}, \bibinfo{author}{G.~Fratesi},
  \bibinfo{author}{R.~Gebauer}, \bibinfo{author}{U.~Gerstmann},
  \bibinfo{author}{C.~Gougoussis}, \bibinfo{author}{A.~Kokalj},
  \bibinfo{author}{M.~Lazzeri}, \bibinfo{author}{L.~Martin-Samos},
  \bibinfo{author}{N.~Marzari}, \bibinfo{author}{F.~Mauri},
  \bibinfo{author}{R.~Mazzarello}, \bibinfo{author}{S.~Paolini},
  \bibinfo{author}{A.~Pasquarello}, \bibinfo{author}{L.~Paulatto},
  \bibinfo{author}{C.~Sbraccia}, \bibinfo{author}{S.~Scandolo},
  \bibinfo{author}{G.~Sclauzero}, \bibinfo{author}{A.~P. Seitsonen},
  \bibinfo{author}{A.~Smogunov}, \bibinfo{author}{P.~Umari},
  \bibinfo{author}{R.~M. Wentzcovitch},
\newblock \bibinfo{title}{{QUANTUM ESPRESSO: a modular and open-source software
  project for quantum simulations of materials}},
\newblock \bibinfo{journal}{J. Phys. Condens. Matter} \bibinfo{volume}{21}
  (\bibinfo{year}{2009}) \bibinfo{pages}{395502}.
\bibitem[{Perdew et~al.(1997)Perdew, Burke, and Ernzerhof}]{Perdew1997}
\bibinfo{author}{J.~P. Perdew}, \bibinfo{author}{K.~Burke},
  \bibinfo{author}{M.~Ernzerhof},
\newblock \bibinfo{title}{{Generalized Gradient Approximation Made Simple
  [Phys. Rev. Lett. 77, 3865 (1996)]}},
\newblock \bibinfo{journal}{Phys. Rev. Lett.} \bibinfo{volume}{78}
  (\bibinfo{year}{1997}) \bibinfo{pages}{1396--1396}.
\bibitem[{Zeng et~al.(2024)Zeng, Wang, Wang, Lin, Gong, Ma, Han, Wang, Dai, and
  Xia}]{Zeng2024}
\bibinfo{author}{X.-Y. Zeng}, \bibinfo{author}{H.~Wang}, \bibinfo{author}{X.-Y.
  Wang}, \bibinfo{author}{J.-F. Lin}, \bibinfo{author}{J.~Gong},
  \bibinfo{author}{X.-P. Ma}, \bibinfo{author}{K.~Han}, \bibinfo{author}{Y.-T.
  Wang}, \bibinfo{author}{Z.-Y. Dai}, \bibinfo{author}{T.-L. Xia},
\newblock \bibinfo{title}{{Magnetic and magnetotransport properties in the
  vanadium-based kagome metals $DyV_6Sn_6$ and $HoV_6Sn_6$}},
\newblock \bibinfo{journal}{Phys. Rev. B} \bibinfo{volume}{109}
  (\bibinfo{year}{2024}) \bibinfo{pages}{104412}.
\bibitem[{Kurumaji et~al.(2024)Kurumaji, Gen, Kitou, and Arima}]{Kurumaji2024}
\bibinfo{author}{T.~Kurumaji}, \bibinfo{author}{M.~Gen},
  \bibinfo{author}{S.~Kitou}, \bibinfo{author}{T.-h. Arima},
\newblock \bibinfo{title}{{Metamagnetism and anomalous magnetotransport
  properties in rare-earth-based polar semimetals $R\mathrm{AuGe}$
  $(R=\mathrm{Dy}, \mathrm{Ho}, \text{and} \mathrm{Gd})$}},
\newblock \bibinfo{journal}{Phys. Rev. B} \bibinfo{volume}{110}
  (\bibinfo{year}{2024}) \bibinfo{pages}{064409}.
\bibitem[{Cadogan and Ryan(2001)}]{Cadogan2001}
\bibinfo{author}{J.~M. Cadogan}, \bibinfo{author}{D.~H. Ryan},
\newblock \bibinfo{title}{{Independent magnetic ordering of the rare-earth (R)
  and Fe sublattices in the $RFe_6Ge_6$ and $RFe_6Sn_6$ series}},
\newblock \bibinfo{journal}{J. Alloy. Compd.} \bibinfo{volume}{326}
  (\bibinfo{year}{2001}) \bibinfo{pages}{166--173}.
\bibitem[{Cadogan et~al.(2001)Cadogan, Suharyana, Ryan, Može, and
  Kockelmann}]{Cadogan2001a}
\bibinfo{author}{J.~M. Cadogan}, \bibinfo{author}{Suharyana},
  \bibinfo{author}{D.~H. Ryan}, \bibinfo{author}{O.~Može},
  \bibinfo{author}{W.~Kockelmann},
\newblock \bibinfo{title}{{Neutron diffraction and Mossbauer study of the
  magnetic structure of $HoFe_6Sn_6$}},
\newblock \bibinfo{journal}{IEEE Transactions on Magnetics}
  \bibinfo{volume}{37} (\bibinfo{year}{2001}) \bibinfo{pages}{2606--2608}.
\bibitem[{Cadogan et~al.(2006)Cadogan, Može, Ryan, Suharyana, and
  Hofmann}]{Cadogan2006}
\bibinfo{author}{J.~M. Cadogan}, \bibinfo{author}{O.~Može},
  \bibinfo{author}{D.~H. Ryan}, \bibinfo{author}{Suharyana},
  \bibinfo{author}{M.~Hofmann},
\newblock \bibinfo{title}{{Magnetic ordering in $DyFe_6Sn_6$}},
\newblock \bibinfo{journal}{Physica B} \bibinfo{volume}{385-386}
  (\bibinfo{year}{2006}) \bibinfo{pages}{317--319}.
\bibitem[{Lv et~al.(2024)Lv, Guo, Qi, Li, Hu, Zheng, Wang, Si, Zhu, Zhao, Han,
  Yu, Xian, Huang, Bao, Lin, Pan, Du, He, Yang, and Gao}]{Lv2024}
\bibinfo{author}{S.~Lv}, \bibinfo{author}{H.~Guo}, \bibinfo{author}{Q.~Qi},
  \bibinfo{author}{Y.~Li}, \bibinfo{author}{G.~Hu}, \bibinfo{author}{Q.~Zheng},
  \bibinfo{author}{R.~Wang}, \bibinfo{author}{N.~Si}, \bibinfo{author}{K.~Zhu},
  \bibinfo{author}{Z.~Zhao}, \bibinfo{author}{Y.~Han}, \bibinfo{author}{W.~Yu},
  \bibinfo{author}{G.~Xian}, \bibinfo{author}{L.~Huang},
  \bibinfo{author}{L.~Bao}, \bibinfo{author}{X.~Lin}, \bibinfo{author}{J.~Pan},
  \bibinfo{author}{S.~Du}, \bibinfo{author}{J.~He}, \bibinfo{author}{H.~Yang},
  \bibinfo{author}{H.~Gao},
\newblock \bibinfo{title}{{Field‐Induced Butterfly‐Like anisotropic
  magnetoresistance in a Kagome semimetal $Co_3In_2S_2$}},
\newblock \bibinfo{journal}{Adv. Funct. Mater.}  (\bibinfo{year}{2024}).
\bibitem[{Singha et~al.(2017)Singha, Pariari, Satpati, and Mandal}]{Singha2017}
\bibinfo{author}{R.~Singha}, \bibinfo{author}{A.~Pariari},
  \bibinfo{author}{B.~Satpati}, \bibinfo{author}{P.~Mandal},
\newblock \bibinfo{title}{{Magnetotransport properties and evidence of a
  topological insulating state in LaSbTe}},
\newblock \bibinfo{journal}{Phys. Rev. B} \bibinfo{volume}{96}
  (\bibinfo{year}{2017}) \bibinfo{pages}{245138}.
\bibitem[{Narayanan et~al.(2015)Narayanan, Watson, Blake, Bruyant, Drigo, Chen,
  Prabhakaran, Yan, Felser, Kong, Canfield, and Coldea}]{Narayanan2015}
\bibinfo{author}{A.~Narayanan}, \bibinfo{author}{M.~D. Watson},
  \bibinfo{author}{S.~F. Blake}, \bibinfo{author}{N.~Bruyant},
  \bibinfo{author}{L.~Drigo}, \bibinfo{author}{Y.~L. Chen},
  \bibinfo{author}{D.~Prabhakaran}, \bibinfo{author}{B.~Yan},
  \bibinfo{author}{C.~Felser}, \bibinfo{author}{T.~Kong},
  \bibinfo{author}{P.~C. Canfield}, \bibinfo{author}{A.~I. Coldea},
\newblock \bibinfo{title}{{Linear Magnetoresistance Caused by Mobility
  Fluctuations in $n$-Doped ${\mathrm{Cd}}_{3}{\mathrm{As}}_{2}$}},
\newblock \bibinfo{journal}{Phys. Rev. Lett.} \bibinfo{volume}{114}
  (\bibinfo{year}{2015}) \bibinfo{pages}{117201}.
\bibitem[{Zhu et~al.(2022)Zhu, Cao, Guo, Li, Chen, Zhu, He, Huang, Dong, Wang,
  Zhai, Ou, Zhu, Lu, Li, Chen, and Pan}]{Zhu2022}
\bibinfo{author}{W.~L. Zhu}, \bibinfo{author}{Y.~Cao}, \bibinfo{author}{P.~J.
  Guo}, \bibinfo{author}{X.~Li}, \bibinfo{author}{Y.~J. Chen},
  \bibinfo{author}{L.~J. Zhu}, \bibinfo{author}{J.~B. He},
  \bibinfo{author}{Y.~F. Huang}, \bibinfo{author}{Q.~X. Dong},
  \bibinfo{author}{Y.~Y. Wang}, \bibinfo{author}{R.~Q. Zhai},
  \bibinfo{author}{Y.~B. Ou}, \bibinfo{author}{G.~Q. Zhu},
  \bibinfo{author}{H.~Y. Lu}, \bibinfo{author}{G.~Li}, \bibinfo{author}{G.~F.
  Chen}, \bibinfo{author}{M.~H. Pan},
\newblock \bibinfo{title}{Linear magnetoresistance induced by mobility
  fluctuations in iodine-intercalated tungsten ditelluride},
\newblock \bibinfo{journal}{Phys. Rev. B} \bibinfo{volume}{105}
  (\bibinfo{year}{2022}) \bibinfo{pages}{125116}.
\bibitem[{Jin et~al.(2012)Jin, Li, Mi, and Bai}]{Jin2012}
\bibinfo{author}{C.~Jin}, \bibinfo{author}{P.~Li}, \bibinfo{author}{W.~B. Mi},
  \bibinfo{author}{H.~L. Bai},
\newblock \bibinfo{title}{{Magnetocrystalline anisotropy-dependent six-fold
  symmetric anisotropic magnetoresistance in epitaxial $Co_xFe_{3- x}O_4$
  films}},
\newblock \bibinfo{journal}{Europhys. Lett.} \bibinfo{volume}{100}
  (\bibinfo{year}{2012}) \bibinfo{pages}{27006}.
\bibitem[{Miao et~al.(2021)Miao, Yang, Jia, Li, Yang, Gao, and Xue}]{Miao2021}
\bibinfo{author}{Y.~Miao}, \bibinfo{author}{D.~Yang}, \bibinfo{author}{L.~Jia},
  \bibinfo{author}{X.~Li}, \bibinfo{author}{S.~Yang}, \bibinfo{author}{C.~Gao},
  \bibinfo{author}{D.~Xue},
\newblock \bibinfo{title}{{Magnetocrystalline anisotropy correlated negative
  anisotropic magnetoresistance in epitaxial $Fe_{30}Co_{70}$ thin films}},
\newblock \bibinfo{journal}{Appl. Phys. Lett.} \bibinfo{volume}{118}
  (\bibinfo{year}{2021}).
\bibitem[{Zhang et~al.(2023)Zhang, Chen, Wang, Zhang, Shi, Zhan, Zhao, Li,
  Zheng, Zhang, Han, Yang, Zhu, Liu, Hu, Shen, Chen, Zhang, Chen, Zhao, and
  Sun}]{Zhang2023}
\bibinfo{author}{J.~Zhang}, \bibinfo{author}{X.~Chen},
  \bibinfo{author}{M.~Wang}, \bibinfo{author}{Q.~Zhang},
  \bibinfo{author}{W.~Shi}, \bibinfo{author}{X.~Zhan},
  \bibinfo{author}{M.~Zhao}, \bibinfo{author}{Z.~Li},
  \bibinfo{author}{J.~Zheng}, \bibinfo{author}{H.~Zhang},
  \bibinfo{author}{F.~Han}, \bibinfo{author}{H.~Yang},
  \bibinfo{author}{T.~Zhu}, \bibinfo{author}{B.~Liu}, \bibinfo{author}{F.~Hu},
  \bibinfo{author}{B.~Shen}, \bibinfo{author}{Y.~Chen},
  \bibinfo{author}{Y.~Zhang}, \bibinfo{author}{Y.~Chen},
  \bibinfo{author}{W.~Zhao}, \bibinfo{author}{J.~Sun},
\newblock \bibinfo{title}{{Proximity‐Induced Fully Ferromagnetic Order with
  Eightfold Magnetic Anisotropy in Heavy Transition Metal Oxide $CaRuO_3$}},
\newblock \bibinfo{journal}{Adv. Funct. Mater.} \bibinfo{volume}{33}
  (\bibinfo{year}{2023}).
\bibitem[{Ramos et~al.(2008)Ramos, Arora, and Shvets}]{Ramos2008}
\bibinfo{author}{R.~Ramos}, \bibinfo{author}{S.~K. Arora},
  \bibinfo{author}{I.~V. Shvets},
\newblock \bibinfo{title}{{Anomalous anisotropic magnetoresistance in epitaxial
  ${\text{Fe}}_{3}{\text{O}}_{4}$ thin films on MgO(001)}},
\newblock \bibinfo{journal}{Phys. Rev. B} \bibinfo{volume}{78}
  (\bibinfo{year}{2008}) \bibinfo{pages}{214402}.
\bibitem[{Li et~al.(2010{\natexlab{a}})Li, Jiang, and Bai}]{Li2010}
\bibinfo{author}{P.~Li}, \bibinfo{author}{E.~Y. Jiang}, \bibinfo{author}{H.~L.
  Bai},
\newblock \bibinfo{title}{{Fourfold symmetric anisotropic magnetoresistance
  based on magnetocrystalline anisotropy and antiphase boundaries in reactive
  sputtered epitaxial $Fe_3O_4$ films}},
\newblock \bibinfo{journal}{Appl. Phys. Lett.} \bibinfo{volume}{96}
  (\bibinfo{year}{2010}{\natexlab{a}}).
\bibitem[{Li et~al.(2010{\natexlab{b}})Li, Jin, Jiang, and Bai}]{Li2010a}
\bibinfo{author}{P.~Li}, \bibinfo{author}{C.~Jin}, \bibinfo{author}{E.~Y.
  Jiang}, \bibinfo{author}{H.~L. Bai},
\newblock \bibinfo{title}{{Origin of the twofold and fourfold symmetric
  anisotropic magnetoresistance in epitaxial $Fe_3O_4$ films}},
\newblock \bibinfo{journal}{J Appl Phys} \bibinfo{volume}{108}
  (\bibinfo{year}{2010}{\natexlab{b}}).
\bibitem[{Li et~al.(2015)Li, Mi, Wang, and Zhang}]{Li2015}
\bibinfo{author}{Z.~Li}, \bibinfo{author}{W.~Mi}, \bibinfo{author}{X.~Wang},
  \bibinfo{author}{X.~Zhang},
\newblock \bibinfo{title}{{Interfacial Exchange Coupling Induced Anomalous
  Anisotropic Magnetoresistance in Epitaxial Fe$_4$N/CoN Bilayers}},
\newblock \bibinfo{journal}{ACS Appl. Mater. Interfaces} \bibinfo{volume}{7}
  (\bibinfo{year}{2015}) \bibinfo{pages}{3840}. \bibinfo{note}{PMID: 25643137}.
\bibitem[{Dai et~al.(2022)Dai, Zhao, Ma, Tang, Qiu, Liu, Yuan, and
  Zhou}]{Dai2022}
\bibinfo{author}{Y.~Dai}, \bibinfo{author}{Y.~W. Zhao},
  \bibinfo{author}{L.~Ma}, \bibinfo{author}{M.~Tang}, \bibinfo{author}{X.~P.
  Qiu}, \bibinfo{author}{Y.~Liu}, \bibinfo{author}{Z.~Yuan},
  \bibinfo{author}{S.~M. Zhou},
\newblock \bibinfo{title}{{Fourfold Anisotropic Magnetoresistance of
  ${\mathrm{L}1}_{0}$ FePt Due to Relaxation Time Anisotropy}},
\newblock \bibinfo{journal}{Phys. Rev. Lett.} \bibinfo{volume}{128}
  (\bibinfo{year}{2022}) \bibinfo{pages}{247202}.
\bibitem[{Zhou et~al.(2020)Zhou, Sun, Xu, Xi, Wang, Li, Feng, Zhang, Xing,
  Zhang, Pan, Meng, Yi, Pi, Xu, and Shi}]{Zhou2020}
\bibinfo{author}{N.~Zhou}, \bibinfo{author}{Y.~Sun}, \bibinfo{author}{C.~Q.
  Xu}, \bibinfo{author}{C.~Y. Xi}, \bibinfo{author}{Z.~S. Wang},
  \bibinfo{author}{B.~Li}, \bibinfo{author}{J.~J. Feng},
  \bibinfo{author}{L.~Zhang}, \bibinfo{author}{X.~Z. Xing},
  \bibinfo{author}{Y.~F. Zhang}, \bibinfo{author}{Y.~Q. Pan},
  \bibinfo{author}{Y.~Meng}, \bibinfo{author}{X.~L. Yi},
  \bibinfo{author}{L.~Pi}, \bibinfo{author}{X.~Xu}, \bibinfo{author}{Z.~Shi},
\newblock \bibinfo{title}{{Quantum oscillations and anomalous angle-dependent
  magnetoresistance in the topological candidate
  ${\mathrm{Ag}}_{3}\mathrm{Sn}$}},
\newblock \bibinfo{journal}{Phys. Rev. B} \bibinfo{volume}{101}
  (\bibinfo{year}{2020}) \bibinfo{pages}{245102}.
\bibitem[{Hu et~al.(2012)Hu, Zhu, Chen, Li, and Wu}]{Hu2012}
\bibinfo{author}{C.~Hu}, \bibinfo{author}{J.~Zhu}, \bibinfo{author}{G.~Chen},
  \bibinfo{author}{J.~Li}, \bibinfo{author}{Y.~Wu},
\newblock \bibinfo{title}{{Direct comparison of anisotropic magnetoresistance
  and planar Hall effect in epitaxial $Fe_3O_4$ thin films}},
\newblock \bibinfo{journal}{Phys. Lett. A} \bibinfo{volume}{376}
  (\bibinfo{year}{2012}) \bibinfo{pages}{3317--3321}.
\bibitem[{Chen et~al.(2023)Chen, Li, Guo, Chen, Zheng, Yu, Lau, Xi, and
  Wang}]{Chen2023}
\bibinfo{author}{J.~Chen}, \bibinfo{author}{H.~Li}, \bibinfo{author}{T.~Guo},
  \bibinfo{author}{P.~Chen}, \bibinfo{author}{D.~Zheng},
  \bibinfo{author}{G.~Yu}, \bibinfo{author}{Y.-C. Lau},
  \bibinfo{author}{X.~Xi}, \bibinfo{author}{W.~Wang},
\newblock \bibinfo{title}{{Anomalous anisotropic magnetoresistance in the
  topological semimetal HoPtBi}},
\newblock \bibinfo{journal}{NPG Asia Mater.} \bibinfo{volume}{15}
  (\bibinfo{year}{2023}).
\bibitem[{Zhu et~al.(2018)Zhu, Fauqué, Behnia, and Fuseya}]{Zhu2018}
\bibinfo{author}{Z.~Zhu}, \bibinfo{author}{B.~Fauqué},
  \bibinfo{author}{K.~Behnia}, \bibinfo{author}{Y.~Fuseya},
\newblock \bibinfo{title}{{Magnetoresistance and valley degree of freedom in
  bulk bismuth}},
\newblock \bibinfo{journal}{J. Phys. Condens. Matter} \bibinfo{volume}{30}
  (\bibinfo{year}{2018}) \bibinfo{pages}{313001}.
\bibitem[{Wang et~al.(2019)Wang, Yang, Ding, You, Xi, Cheng, Shi, Cao, Luo,
  Zhu, Dai, Tian, and Li}]{Wang2019}
\bibinfo{author}{J.~Wang}, \bibinfo{author}{H.~Yang},
  \bibinfo{author}{L.~Ding}, \bibinfo{author}{W.~You}, \bibinfo{author}{C.~Xi},
  \bibinfo{author}{J.~Cheng}, \bibinfo{author}{Z.~Shi},
  \bibinfo{author}{C.~Cao}, \bibinfo{author}{Y.~Luo}, \bibinfo{author}{Z.~Zhu},
  \bibinfo{author}{J.~Dai}, \bibinfo{author}{M.~Tian}, \bibinfo{author}{Y.~Li},
\newblock \bibinfo{title}{{Angle-dependent magnetoresistance and its
  implications for Lifshitz transition in $W_2As_3$}},
\newblock \bibinfo{journal}{npj Qantum Materials} \bibinfo{volume}{4}
  (\bibinfo{year}{2019}).
\bibitem[{Zhang et~al.(2019)Zhang, Wu, Liu, and Yazyev}]{Zhang2019}
\bibinfo{author}{S.~Zhang}, \bibinfo{author}{Q.~Wu}, \bibinfo{author}{Y.~Liu},
  \bibinfo{author}{O.~V. Yazyev},
\newblock \bibinfo{title}{{Magnetoresistance from Fermi surface topology}},
\newblock \bibinfo{journal}{Phys. Rev. B} \bibinfo{volume}{99}
  (\bibinfo{year}{2019}) \bibinfo{pages}{035142}.
\bibitem[{Weiland et~al.(2020)Weiland, Eddy, McCandless, Hodovanets, Paglione,
  and Chan}]{Weiland2020}
\bibinfo{author}{A.~Weiland}, \bibinfo{author}{L.~J. Eddy},
  \bibinfo{author}{G.~T. McCandless}, \bibinfo{author}{H.~Hodovanets},
  \bibinfo{author}{J.~Paglione}, \bibinfo{author}{J.~Y. Chan},
\newblock \bibinfo{title}{{Refine intervention: Characterizing disordered
  $Yb_{0.5}Co_3Ge_3$}},
\newblock \bibinfo{journal}{Cryst. Growth Des.} \bibinfo{volume}{20}
  (\bibinfo{year}{2020}) \bibinfo{pages}{6715--6721}.
\bibitem[{Huang et~al.(2023)Huang, Cui, Huang, Huo, Liu, Li, Liang, Chen, Sun,
  Shen, Zhang, and Wang}]{Huang2023}
\bibinfo{author}{X.~Huang}, \bibinfo{author}{Z.~Cui},
  \bibinfo{author}{C.~Huang}, \bibinfo{author}{M.~Huo},
  \bibinfo{author}{H.~Liu}, \bibinfo{author}{J.~Li},
  \bibinfo{author}{F.~Liang}, \bibinfo{author}{L.~Chen},
  \bibinfo{author}{H.~Sun}, \bibinfo{author}{B.~Shen},
  \bibinfo{author}{Y.~Zhang}, \bibinfo{author}{M.~Wang},
\newblock \bibinfo{title}{{Anisotropic magnetism and electronic properties of
  the kagome metal $SmV_6Sn_6$}},
\newblock \bibinfo{journal}{Phys. Rev. Mater.} \bibinfo{volume}{7}
  (\bibinfo{year}{2023}) \bibinfo{pages}{054403}.
\bibitem[{Guo et~al.(2023)Guo, Ye, Guan, and Jia}]{Guo2023}
\bibinfo{author}{K.~Guo}, \bibinfo{author}{J.~Ye}, \bibinfo{author}{S.~Guan},
  \bibinfo{author}{S.~Jia},
\newblock \bibinfo{title}{{Triangular Kondo lattice in
  ${\mathrm{YbV}}_{6}{\mathrm{Sn}}_{6}$ and its quantum critical behavior in a
  magnetic field}},
\newblock \bibinfo{journal}{Phys. Rev. B} \bibinfo{volume}{107}
  (\bibinfo{year}{2023}) \bibinfo{pages}{205151}.
\bibitem[{Xu et~al.(2023)Xu, Yin, Qu, and Jia}]{Xu2023}
\bibinfo{author}{X.~Xu}, \bibinfo{author}{J.~Yin}, \bibinfo{author}{Z.~Qu},
  \bibinfo{author}{S.~Jia},
\newblock \bibinfo{title}{{Quantum interactions in topological R166 kagome
  magnet}},
\newblock \bibinfo{journal}{Rep. Prog. Phys.} \bibinfo{volume}{86}
  (\bibinfo{year}{2023}) \bibinfo{pages}{114502}.
\bibitem[{Zheng et~al.(2016)Zheng, Lu, Zhu, Ning, Han, Zhang, Zhang, Xi, Yang,
  Du, Yang, Zhang, and Tian}]{Zheng2016}
\bibinfo{author}{G.~Zheng}, \bibinfo{author}{J.~Lu}, \bibinfo{author}{X.~Zhu},
  \bibinfo{author}{W.~Ning}, \bibinfo{author}{Y.~Han},
  \bibinfo{author}{H.~Zhang}, \bibinfo{author}{J.~Zhang},
  \bibinfo{author}{C.~Xi}, \bibinfo{author}{J.~Yang}, \bibinfo{author}{H.~Du},
  \bibinfo{author}{K.~Yang}, \bibinfo{author}{Y.~Zhang},
  \bibinfo{author}{M.~Tian},
\newblock \bibinfo{title}{{Transport evidence for the three-dimensional Dirac
  semimetal phase in $\mathrm{ZrT}{\mathrm{e}}_{5}$}},
\newblock \bibinfo{journal}{Phys. Rev. B} \bibinfo{volume}{93}
  (\bibinfo{year}{2016}) \bibinfo{pages}{115414}.
\bibitem[{Yu et~al.(2017)Yu, Wang, Lou, Guo, Xu, Liu, Wang, and Xia}]{Yu2017}
\bibinfo{author}{Q.-H. Yu}, \bibinfo{author}{Y.-Y. Wang},
  \bibinfo{author}{R.~Lou}, \bibinfo{author}{P.-J. Guo},
  \bibinfo{author}{S.~Xu}, \bibinfo{author}{K.~Liu}, \bibinfo{author}{S.~Wang},
  \bibinfo{author}{T.-L. Xia},
\newblock \bibinfo{title}{{Magnetoresistance and Shubnikov-de Haas oscillation
  in YSb}},
\newblock \bibinfo{journal}{EPL (Europhysics Letters)} \bibinfo{volume}{119}
  (\bibinfo{year}{2017}) \bibinfo{pages}{17002}.
\bibitem[{Wang et~al.(2018)Wang, Guo, Sun, Li, Liu, Lu, and Lei}]{Wang2018a}
\bibinfo{author}{Q.~Wang}, \bibinfo{author}{P.-J. Guo},
  \bibinfo{author}{S.~Sun}, \bibinfo{author}{C.~Li}, \bibinfo{author}{K.~Liu},
  \bibinfo{author}{Z.-Y. Lu}, \bibinfo{author}{H.~Lei},
\newblock \bibinfo{title}{{Extremely large magnetoresistance and high-density
  Dirac-like fermions in ${\mathrm{ZrB}}_{2}$}},
\newblock \bibinfo{journal}{Phys Rev B} \bibinfo{volume}{97}
  (\bibinfo{year}{2018}) \bibinfo{pages}{205105}.
\bibitem[{Lv et~al.(2018)Lv, Li, Zhang, Pang, Chen, Cao, Zhang, Lin, Chen, Yao,
  Zhou, Zhang, Lu, Tian, and Chen}]{Lv2018}
\bibinfo{author}{Y.-Y. Lv}, \bibinfo{author}{X.~Li},
  \bibinfo{author}{J.~Zhang}, \bibinfo{author}{B.~Pang}, \bibinfo{author}{S.-S.
  Chen}, \bibinfo{author}{L.~Cao}, \bibinfo{author}{B.-B. Zhang},
  \bibinfo{author}{D.~Lin}, \bibinfo{author}{Y.~B. Chen},
  \bibinfo{author}{S.-H. Yao}, \bibinfo{author}{J.~Zhou},
  \bibinfo{author}{S.-T. Zhang}, \bibinfo{author}{M.-H. Lu},
  \bibinfo{author}{M.~Tian}, \bibinfo{author}{Y.-F. Chen},
\newblock \bibinfo{title}{{Mobility-controlled extremely large
  magnetoresistance in perfect electron-hole compensated $\alpha -WP_2$
  crystals}},
\newblock \bibinfo{journal}{Phys. Rev. B} \bibinfo{volume}{97}
  (\bibinfo{year}{2018}) \bibinfo{pages}{245151}.

\end{thebibliography}

\section{Supplementary Information}
\setcounter{figure}{0}
\renewcommand{\figurename}{Fig.~S}
\begin{figure*}[t]
\center
	\includegraphics[width=0.75\linewidth]{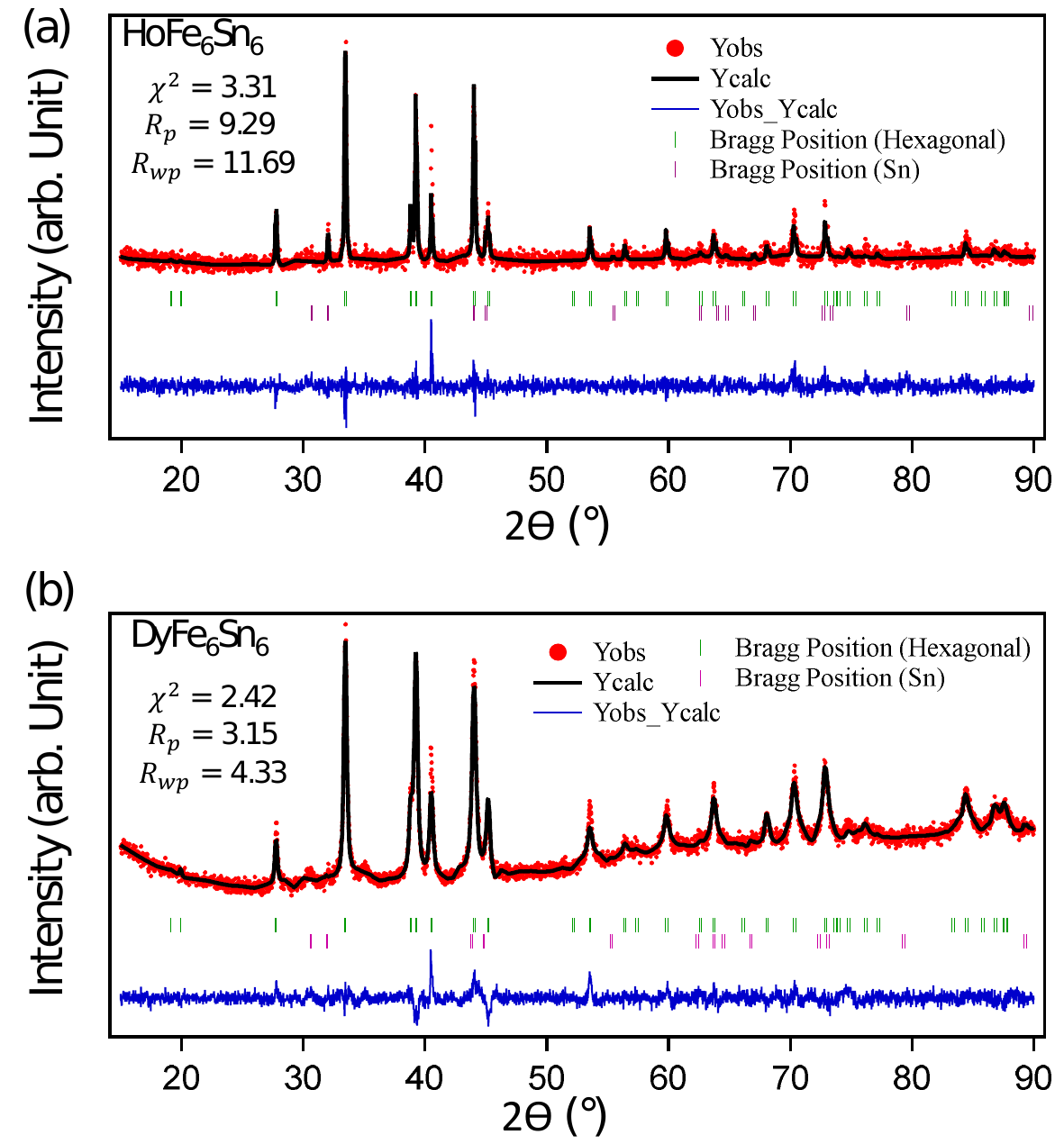}
	\caption{(a) Powder X-ray diffraction pattern of the crushed single crystals of $HoFe_6Sn_6$ (a) and  $DyFe_6Sn_6$ (b), overlapped with Rietveld refinement. Since large number of single crystals were crushed for powder XRD, some amount of Sn flux also got mixed with the powders, resulting in as impurity peaks.}
	\label{figS1}
\end{figure*}

\begin{figure*}[!ht]
\center
	\includegraphics[width=0.85\linewidth]{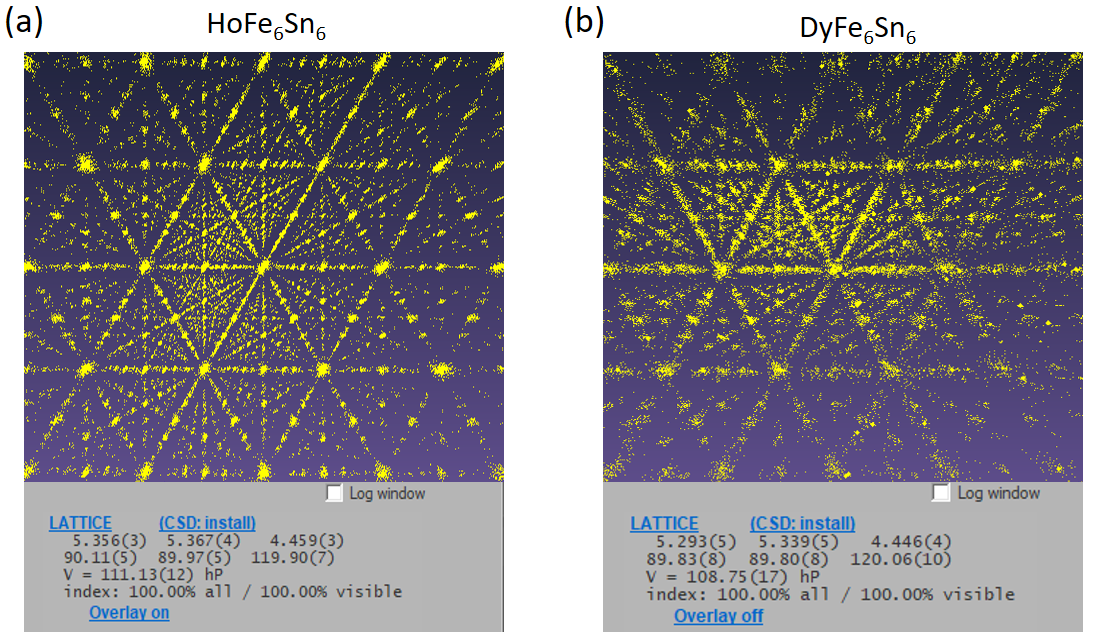}
	\caption{ Gnomonic projection of single crystal XRD (SCXRD) pattern using the Ewald's explorer-reciprocal space option in Crysalispro software of (a) $HoFe_6Sn_6$ and  (b) $DyFe_6Sn_6$. The SCXRD data show perfect hexagonal pattern and the obtained lattice parameters are mentioned in the bottom of respective panels,  confirming the Hexagonal crystal structure of the studied systems.}
	\label{figS2}
\end{figure*}


\begin{figure*}[t]
\center
	\includegraphics[width=0.75\linewidth]{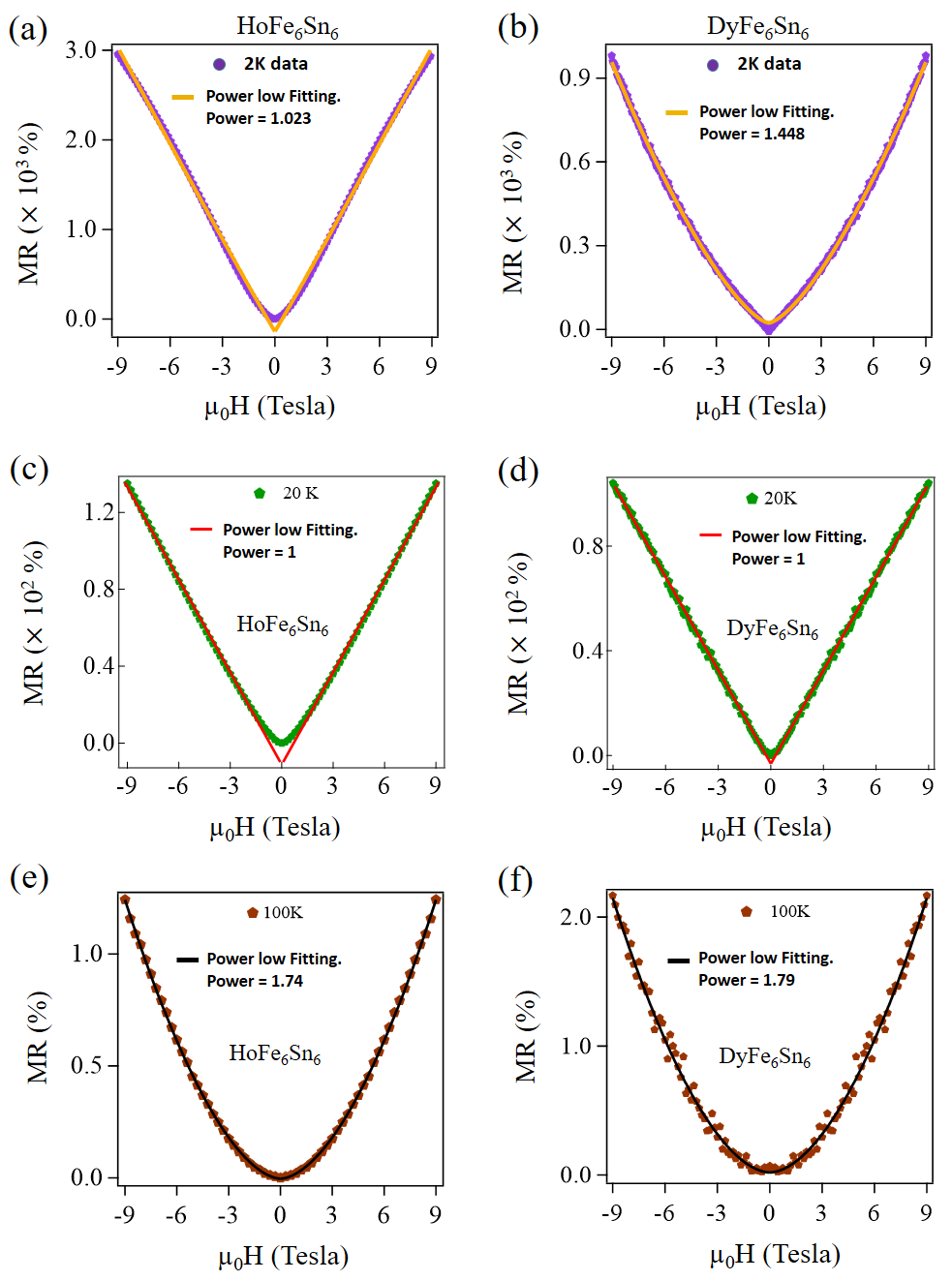}
	\caption{(a)-(f) Field-dependent magnetoresistance (MR) fitted with power law (MR $\varpropto$ $B^m$) at different temperatures for $HoFe_6Sn_6$ and $DyFe_6Sn_6$.}
	\label{figS3}
\end{figure*}

\begin{figure*}[ht]
\center
\includegraphics[width=0.70\linewidth, clip=true]{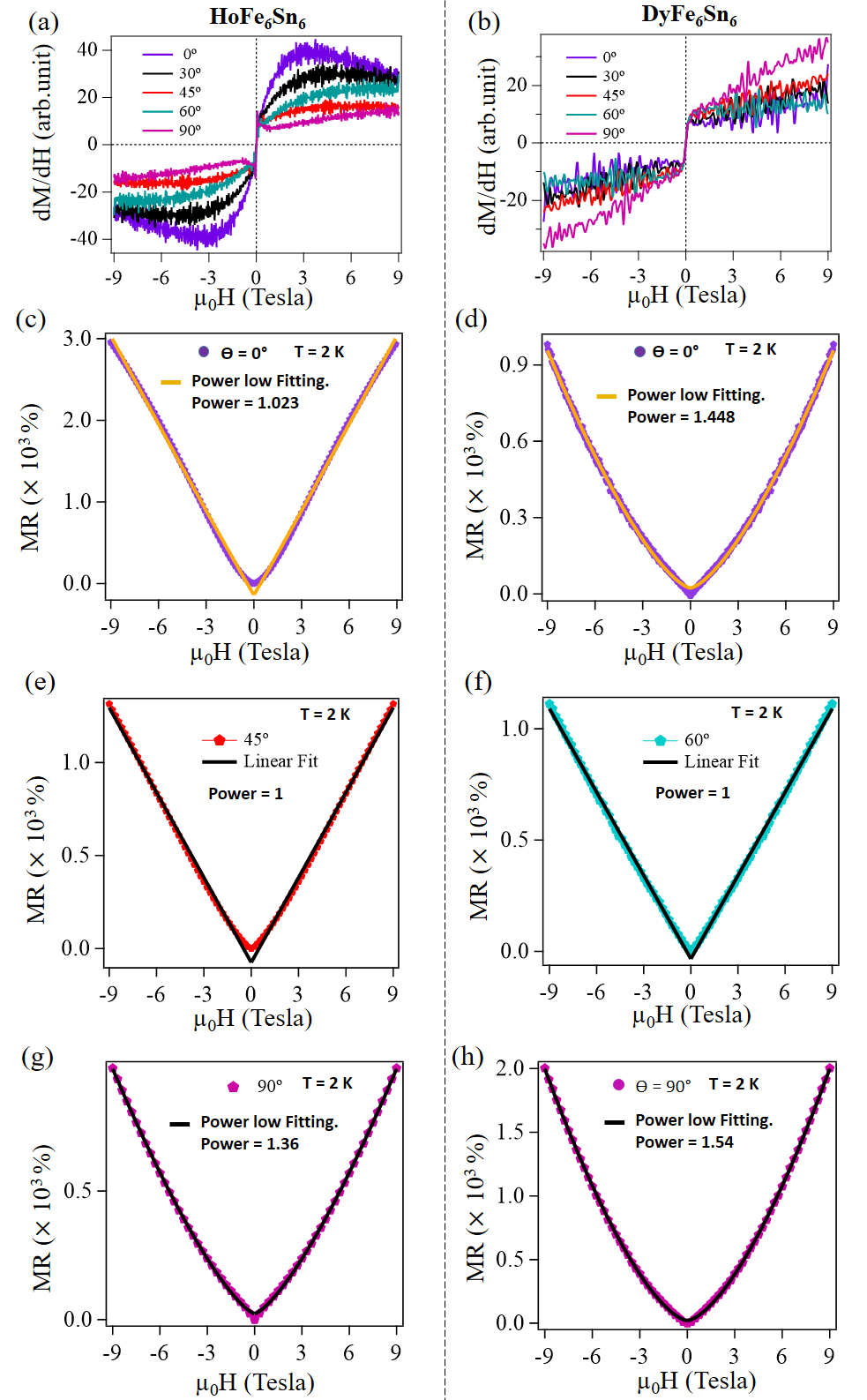}
\caption{ First derivative of MR with respect to field at different field angles measured at 2 K for (a) $HoFe_6Sn_6$ and (b) $DyFe_6Sn_6$. (c), (e), (g), and (d), (f), (h)  are field-dependent MR fitted with power law (MR $\varpropto$ $B^m$ at different field angles measured at 2 K for $HoFe_6Sn_6$ and $DyFe_6Sn_6$ respectively.}
\label{figS4}
\end{figure*}




\begin{figure*}[ht]
\center
\includegraphics[width=1\linewidth, clip=true]{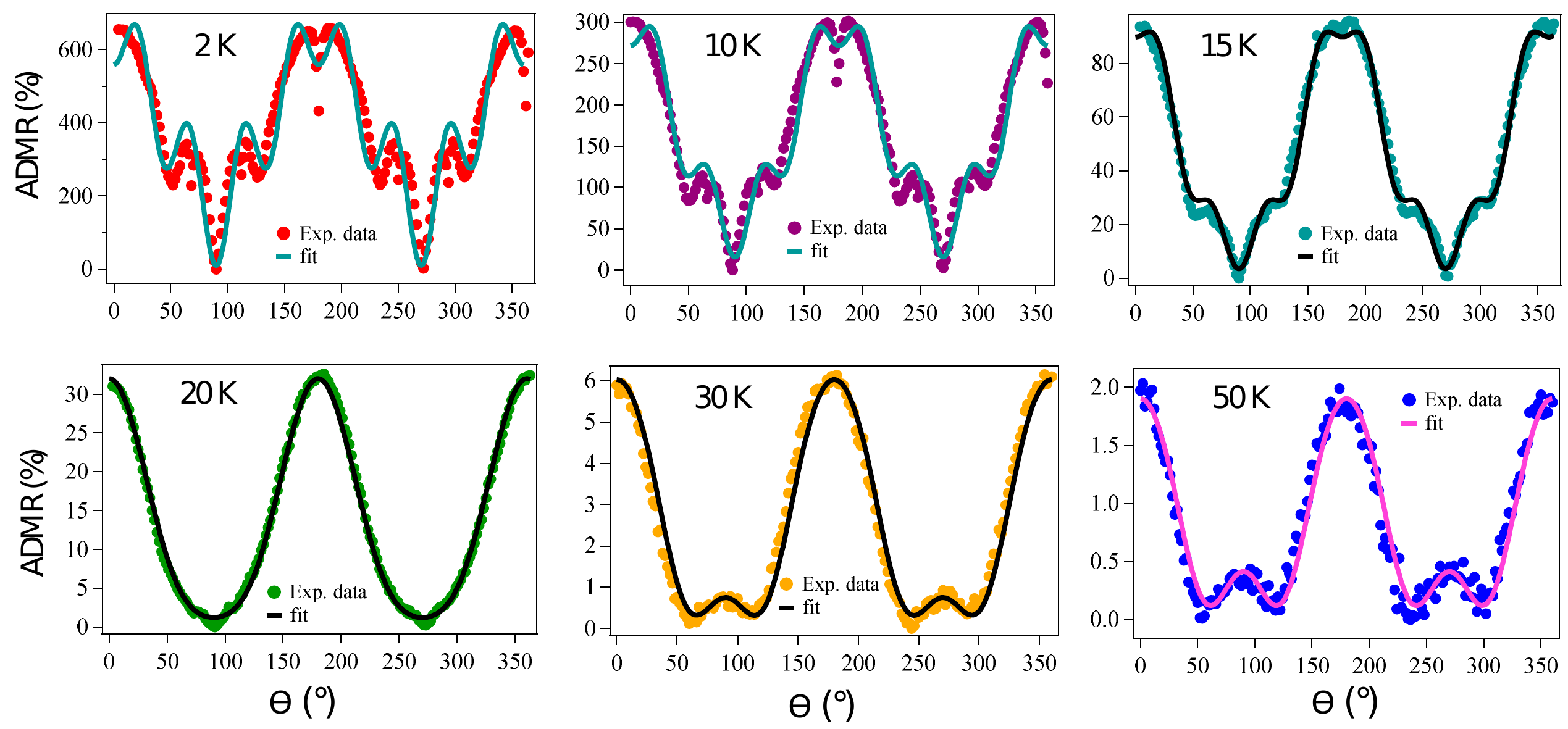}
\caption{ Angle-dependent magnetoresistance (ADMR) measured at various temperatures under the magnetic field of 9 T for $HoFe_6Sn_6$. The solid lines show fits of ADMR.}
\label{figS5}
\end{figure*}

\begin{figure*}[ht]
\center
\includegraphics[width=1\linewidth, clip=true]{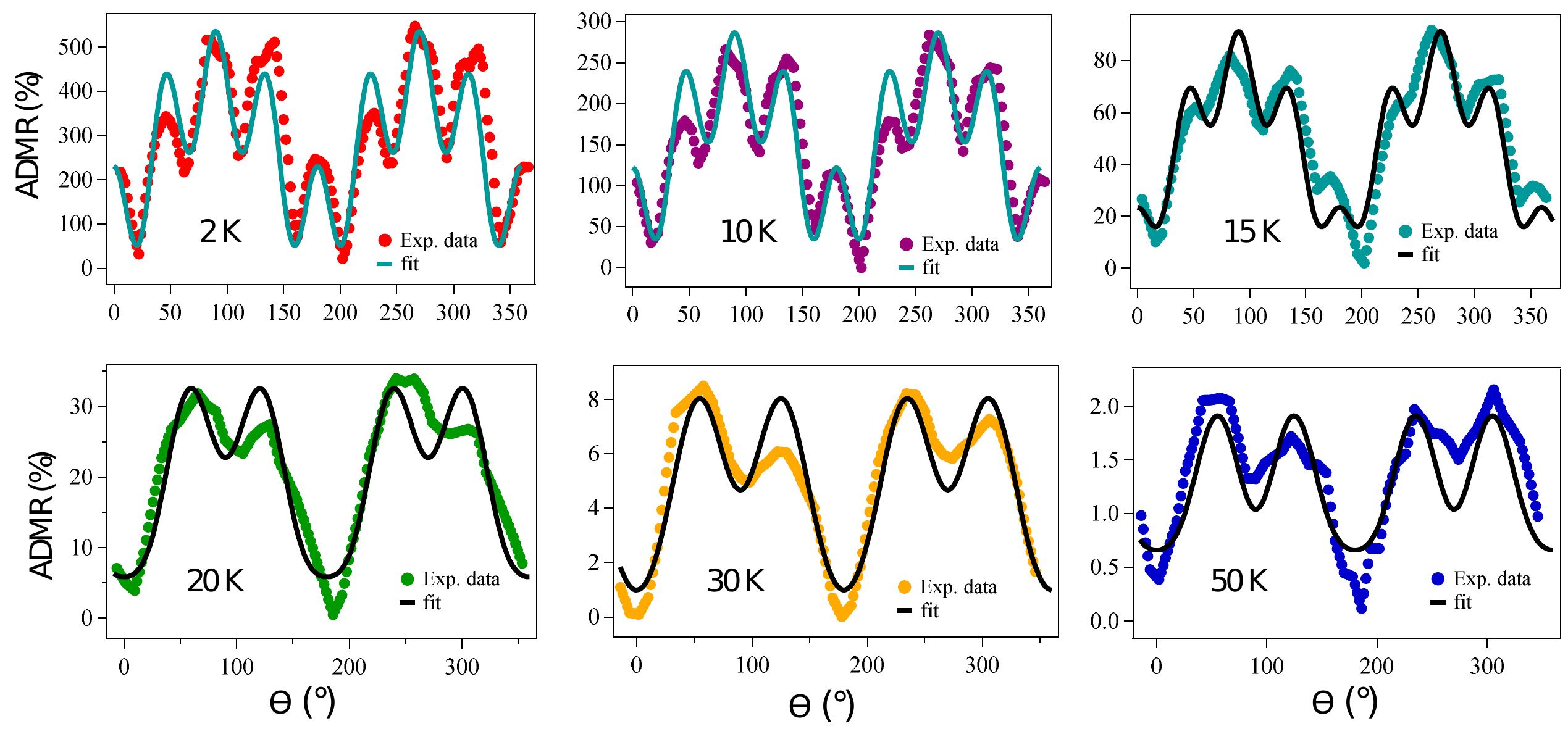}
\caption{ Angle-dependent magnetoresistance (ADMR) measured at various temperatures under the magnetic field of 9 T for $DyFe_6Sn_6$. The solid lines show fits of ADMR.}
\label{figS6}
\end{figure*}

\begin{figure*}[ht]
\center
\includegraphics[width=1\linewidth, clip=true]{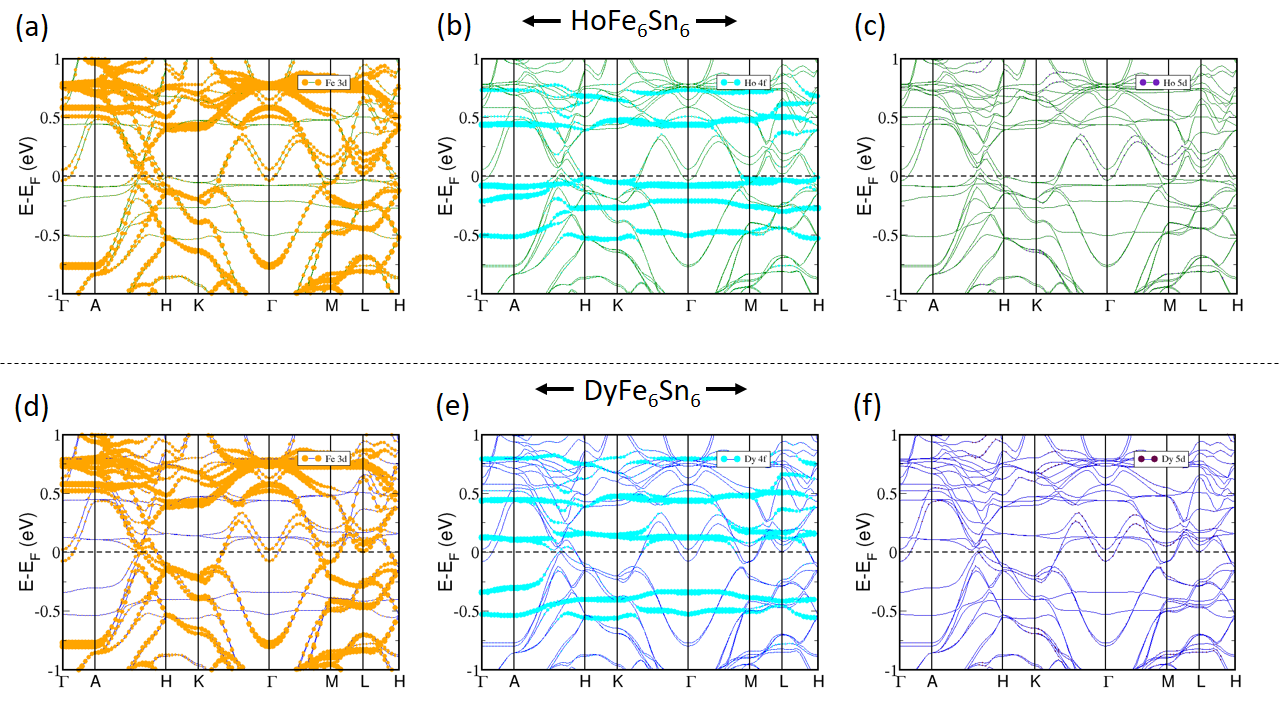}
\caption{ Band structure of  $HoFe_6Sn_6$ for (a) Fe $3d$, (b) Ho $4f$, and (c) Ho $5d$ orbitals. Band structure of  $DyFe_6Sn_6$ for (d) Fe $3d$, (e) Dy $4f$, and (f) Dy 5d orbitals.}
\label{figS7}
\end{figure*}

\begin{figure*}[ht]
\center
\includegraphics[width=1\linewidth, clip=true]{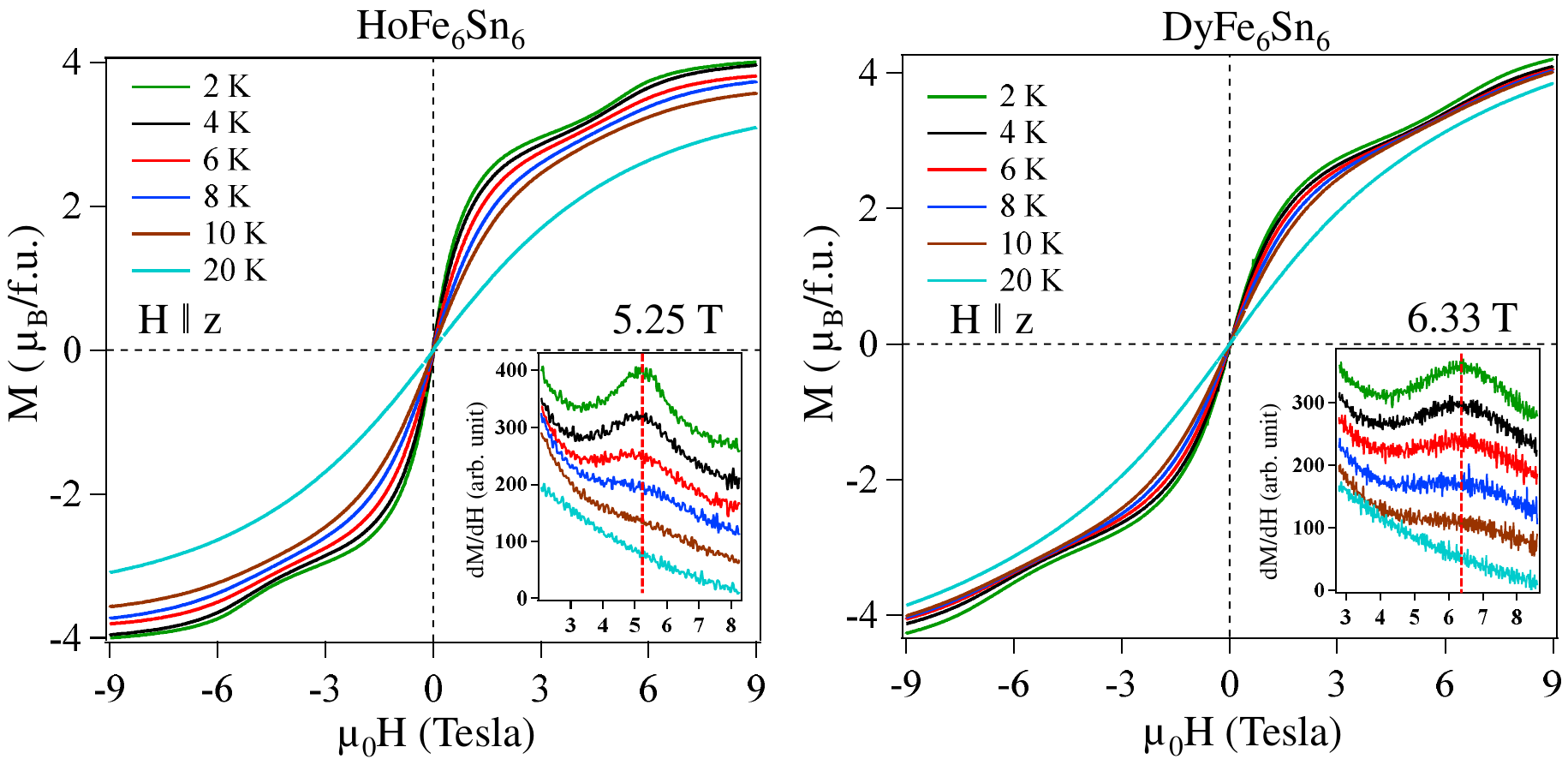}
\caption{ Isothermal magnetization $M(H)$ data taken at various temperatures for HoFe$_6$Sn$_6$ (a) and DyFe$_6$Sn$_6$ (b). First derivatives (d$M$/d$H$) are shown in the insets. Vertical dashed line in the insets suggest the field position of metamagnetic state.}
\label{figS8}
\end{figure*}

\end{document}